# Gas Jet Morphology and the Very Rapidly Increasing Rotation Period of Comet 41P/Tuttle-Giacobini-Kresák

David G. Schleicher[1], Matthew M. Knight[2], Nora L. Eisner[1,2,3], & Audrey Thirouin[1]

[1]Lowell Observatory, 1400 W. Mars Hill Rd., Flagstaff, AZ 86001, USA; dgs@lowell.edu

[2]Dept. of Astronomy, University of Maryland, College Park, MD 20742, USA

[3]Dept. of Physics and Astronomy, University of Sheffield, Sheffield, S3 7RH, UK

Suggested title for running header: The Rapidly Increasing Rotation Period of Comet 41P/Tuttle-Giacobini-Kresák

# ABSTRACT

We present results from our 47-night imaging campaign of Comet 41P/Tuttle-Giacobini-Kresák conducted from Lowell Observatory between 2017 February 16 and July 2. Coma morphology revealed gas jets, whose appearance and motion as a function of time yielded the rotation period and other properties. All narrowband CN images exhibited either one or two jets; one jet appeared as a partial face-on spiral with clockwise rotation while the second jet evolved from a side-on corkscrew, through face-on, and finally corkscrew again, with only a slow evolution throughout the apparition due to progressive viewing geometry changes. A total of 78 period determinations were made over a 7-week interval, yielding a smooth and accelerating rotation period starting at 24 hr (March 21&22) and passing 48 hr on April 28. While this is by far the fastest rate of change ever measured for a comet nucleus, the torque required is readily within what can exist given likely properties of the nucleus. If the torque remained constant, we estimate that the nucleus could have stopped rotating and/or began to tumble as soon as only two months following perihelion, and will certainly reach this stage by early in the next apparition. Working backwards in time, Tuttle-Giacobini-Kresák would have been rotating near its rotational break-up velocity 3-4 orbits earlier, suggesting that its extreme 7-magnitude outburst observed in 2001 might have been caused by a partial fragmentation at that time, as might the pair of 1973 8-magnitude outbursts if there had been an earlier spin-down and spin-up cycle.

# 1. INTRODUCTION

Historically, Comet 41P/Tuttle-Giacobini-Kresák (T-G-K) has been quite interesting for several reasons. Following its initial discovery in 1858 by H. Tuttle, it was thought for many years to have had a long-period orbit. Eventually determined to be a Jupiter-family object, multiple predictions for its recovery were made but each attempt failed before it was independently re-discovered by M. Giacobini in 1907. The two discoveries were subsequently linked but new predictions also failed to result in its recovery; eventually it was again re-discovered, this time in 1951 by L. Kresák, and then recovered at the favorable apparitions of 1962 and 1973 by E. Roemer. At this latter apparition, T-G-K experienced two large (>8 mag) outbursts, and in 2001 it had another major (>7 mag) outburst while evidence also exists for smaller outbursts at other perihelion passages (cf. Kronk 2002, 2007, 2009; Kronk & Meyer 2010; S. Yoshida[1]). Outbursts not withstanding, T-G-K is a relatively inactive comet simply based on the large number of apparitions at which it was *not* detected – since 1858, perihelion has only varied between 1.045 AU (current) and 1.23 AU, and the minimum geocentric distance was less than 1 AU at three other apparitions prior to 1951 at which it was not discovered – suggesting either a relatively small nuclear size and/or a small active fraction.

With these characteristics in mind, the unusually good observing circumstances during the 2017 apparition – better than any previous one – made for an exceptional opportunity to investigate T-G-K in detail, especially as relatively little information regarding this comet's properties was known. In particular, perigee (0.142 AU) occurred on April 1, only 11 days before perihelion (1.045 AU) and nearly at opposition. Moreover, the comet remained within 0.15 AU of Earth for 21 days and within 0.20 AU for 60 days, permitting both excellent spatial and temporal coverage. Predicted to reach $8^{th}$ magnitude, T-G-K was observed by numerous investigators using a wide variety of techniques. Most relevant here, the first report of a measurement of the comet's rotation period was made by Farnham et al. (2017) based on CN jet morphology primarily obtained in early March. Radar measurements from Arecibo by Howell et al. (2017) also provided constraints on the nucleus's properties, suggesting shorter rotation periods than found here.

Our own observing campaign emphasized physical studies using broad and narrowband filter imaging, in particular coma morphology, supplemented with monthly narrowband photometry to measure production rates and basic chemical composition. As expected, light from the nucleus itself could not be separated from that from the inner coma and a nuclear lightcurve could not be measured. Our primary goal was therefore detection and analyses of gas and/or dust jets, similar to what we have done on several other comets, including Hyakutake (C/1996 B2; Schleicher & Woodney 2003), Lulin (C/2007 N3; Bair et al. 2018), 103P/Hartley 2 (Knight & Schleicher 2011, 2013), and 10P/Tempel 2 (Knight et al. 2012). Going into the apparition, we considered it very likely that T-G-K would have one or more distinctive jets since jets require isolated source regions and the comet's relatively low activity level suggested that much of the surface of the nucleus was inert. In particular, the excellent observing circumstances meant that we would be able to observe any jets from a rapidly changing wide range of viewing geometries, thereby allowing us to ultimately create an unambiguous 3-D model of the coma that would yield fundamental properties of the nucleus including the sidereal rotation period, the pole orientation, and the location and size of the source regions for the jets. We also hoped to measure in detail the aftermath of a strong outburst, to better characterize outburst morphology similar to our analyses of the Deep Impact ejecta off of 9P/Tempel 1 (Schleicher et al. 2006) or the eventual decay in production rates following an outburst such as we performed for Comet 17P/Holmes (Schleicher 2009); however, no outbursts of T-G-K were reported during the 2017 apparition.

---

[1] http://www.aerith.net/comet/catalog/0041P

Despite the lack of a major outburst, T-G-K proved more interesting than anyone could have predicted for just one reason – its apparent rotation period changed at a far greater rate than has ever been measured for any object before. As noted, the first reported period was by Farnham et al. (2017), who found a value of 19.9±0.15 hr based on CN jet imaging between March 6 and 9 and supplemented by an image on February 24. They also eliminated a number of potential alias periods and claimed that their solution was indeed the rotation period. We were, therefore, quite surprised to see nearly identical morphology on three successive nights beginning on March 19 *at the same time each night,* directly implying an ~24 hr period, i.e. a 20% increase. Subsequent nights implied a somewhat longer value, and by a week later (March 26 & 27), multiple image pairs from the two nights indicated that the period had further lengthened to nearly 27 hr (Knight et al. 2017). As we will show, the period progressively and systematically lengthened through the apparition at an increasing rate, eventually reaching 48+ hr by the end of April (Schleicher et al. 2017). An independent period determination of 42±1 hr was made by Bodewits et al. (2017a) from a partial lightcurve derived from the *Swift* satellite on May 7-9; this was later revised to 50 hr (Bodewits et al. 2017b), and then to 53±7 hr (Bodewits et al. 2018). Most recently, Moulane et al. (2018) have reported rotation periods of 30±5 hr on March 31 and 50±10 hr on April 27 based on the rate of rotational motion of jets within individual nights.

In this paper we present CN narrowband imaging obtained on a total of 46 nights between 2017 February 16 and July 2, along with imaging of other species obtained on a less frequent basis. The evolving jet morphology is discussed and differences among species examined. Our period determination methodologies are presented along with the detailed results. Although the jet modeling results, along with gas and dust production rates, will be deferred until the jet modeling is completed, we utilize preliminary model results here to assist with some of the period determinations by helping to confirm our identifications of which jet is which from one night to another.

## 2. OBSERVATIONS, REDUCTIONS, AND METHODOLOGIES

*2.1 Overview, Instrumentation, and Observations*

All of the images reported on here were obtained with one of three Lowell Observatory telescopes: the 4.3-m Discovery Channel Telescope (DCT), the John S. Hall 42-in (1.1-m), and the robotic 31-in (0.8-m). While the DCT observations were mostly snap-shots obtained on nights designated for other projects, they provided the best signal-to-noise ratios and extended late into the apparition when T-G-K was otherwise too faint for the smaller telescopes. These images were obtained with the Large Monolithic Imager (LMI), having a 6.1K×6.1K e2v CCD with a 12.3 arcmin field of view. A pixel scale of 0.24 arcsec results from on-chip 2×2 binning. The 42-in has a 4K×4K e2v CCD231-84 chip and a 2×2 binned pixel scale of 0.74 arcsec, while the 31-in has a 2K×2K e2v CCD42-40 chip and a pixel scale of 0.46 arcsec. Matching sets of narrowband HB comet filters (Farnham et al. 2000), differing only in their physical sizes, are used at each telescope, supplemented with a broadband *R* filter for improved S/N of the dust. Snapshots usually consisted of only the *R* and CN filters, while the number of other narrowband filters utilized (OH, CN, $C_3$, and $C_2$ emission bands, along with blue and green continuum points, varied with both the brightness of T-G-K and the observing window each night. While we had scheduled two primary runs, in late March and late April, our early discovery of the rapidly changing rotation period significantly altered our plans and we added as many nights of imaging as possible. Ultimately, successful imaging was obtained on a total of 47 nights (46 with the CN filter), six of which had observations obtained at multiple telescopes. The complete list is given in Table 1, along with other pertinent parameters. Individual exposures were generally 30-60 s for *R*, and 120-300 s for CN. Note that images were usually obtained in sets of three (with an *R*-band set followed by one or more narrowband sets) that were later median combined both to improve S/N and so that individual frames could be removed as needed due to cosmic ray strikes, badly placed background stars, or varying thickness of clouds.

[TABLE 1 HERE: OBSERVATIONAL CIRCUMSTANCES; FILTERS]

*2.2 CCD Reductions and Analyses*

Standard techniques were followed for bias removal and flat fielding. Photometric calibrations were only applied on photometric nights for performing continuum subtraction of the $C_2$ and $C_3$ images, as dust can be a significant contributor within these filters. Fortunately, the dust contribution within the CN and OH filters is small and does not affect the bulk gas morphology. Thus, CN imaging for morphological studies did not require photometric conditions nor absolute calibrations.

Centroids were readily determined using a two-dimensional Gaussian fit of the inner-most coma, both for continuum image subtraction when needed, and for applying our standard enhancement technique of computing and removing the mean radial profile from each image (cf. Schleicher & Farnham 2004; Knight & Schleicher 2015). In particular, this method is quite benign and does not introduce structures nor significantly alter the position of structures within the coma beyond several pixels from the center, thus avoiding possible centroiding uncertainties. A small amount of Gaussian smoothing was applied to lower S/N images. For intercomparisons of morphology, images were trimmed to a common physical scale at the comet of 30,000 km.

3. IMAGING

*3.1 Coma Morphology Overview*

On the vast majority of nights for which we have images of T-G-K the comet was not the primary target. Rather, one or a few sets were interleaved with other targets and projects by various investigators, especially very early and then late in the apparition. As noted, in such cases, a brief set of *R*-band images was immediately followed by a set of CN images. We, therefore, first present and discuss the overall appearance of T-G-K in these two filters throughout the 4½ month interval, with representative images given in Figure 1. All CN images exhibit either one or two gas jets, and on most nights both jets appear as partial spirals with a clockwise rotation. Jets are strongest in the sunward hemisphere and become weaker or even disappear as the source regions rotate into presumed night; the solar illumination function can be determined from the jet modeling and the derived pole orientation and source latitudes. Also, the long-term trends in the shapes and locations of the jets can largely be attributed to changes in our viewing geometries.

[FIGURE 1 HERE: Pair-wise (CN and R) representative images throughout apparition – 5 sets]

As we have often found to be the case in other comets (e.g., 103P/Hartley 2 (Knight & Schleicher 2013); C/Lulin (2007 N3; Bair et al. 2018)), there was little similarity in the dust morphology to that of the gas species in T-G-K and, more importantly for this paper, little to no evidence of significant rotational variations. Rather, a strong dust tail is evident throughout, nearly in the anti-solar direction, and often a weaker, linear feature somewhat offset from the sunward direction. Since rotational variations are so much easier to detect in the CN jets, we defer more detailed investigation of the dust to the next planned paper.

*3.2 Morphology Inter-comparisons Among Gas Species*

In addition to our extensive coverage of CN, we also imaged T-G-K using filters isolating $C_2$, $C_3$, and OH emission during our two primary observing runs in late March and late April; note that the comet was too faint to obtain NH imaging. As expected, the raw $C_2$ and $C_3$ images exhibited structures matching those in both the CN and *R*-band images, but once the $C_2$ and $C_3$ images were flux calibrated and continuum subtracted, they generally looked quite similar to those of CN, but with the expected variations due to parentage and lifetimes (see Figure 2). Specifically, $C_3$ jets do not extend as far due to its and its parent's relatively short lifetimes, while the $C_2$ jets appear somewhat broader than CN due to having multiple parents as well as grandparents. However, when both CN jets are evident at a particular time (such as the middle and bottom rows in Figure 2), the relative intensities between the two CN jets appears to be more similar than for the $C_3$ and $C_2$ counterparts. In fact, it is difficult to discern the fainter jet in these other species, especially in $C_2$.

One possibility is that the relative abundances of $C_3$ and $C_2$ to CN differs between the two source regions, since our modeling indicates that it is the same jet that dominates in each case.

[FIGURE 2 HERE: OH/CN/$C_3$/$C_2$ on representative nights (2 rotational phases in March; 1 in April]

In contrast, the OH images look very different from the other species. First, there is one and only one feature evident in any OH image, and on any particular night it shows little or no rotational motion. Note that the concentric rings in the OH enhanced images are an artifact of the very low signal-to-noise coupled with the removal of the bulk radial profile in the enhancement process. While the observing interval each night was shorter than that for CN due to airmass constraints, the CN jets clearly moved within the same intervals. In fact, when examining the images from March 22, during the interval from 5 UT to 9 UT, one CN jet had essentially turned off and the second jet became dominant with a lagging orientation by more than 90°, while the OH looks nearly unchanged during the same interval. Overall, between March 22 and 27, the only clear change in OH is that the feature progressed from being centered towards the west-southwest to the west-northwest. During the second run, OH was only imaged on three nights, looking very similar on April 21 and 22 (pointing towards the north-northwest), but clearly having shifted clockwise on April 24 (north-west).

Our first conclusion is obvious – the OH feature does not track those of the three minor species we observed and its parent, water, is unlikely to have been primarily emitted by either of the two source regions associated with those jets. The lack of obvious rotational motion suggests either very slowly moving molecules (perhaps vaporizing from icy grains) or, we think more likely, the OH molecules are arising from a different source region perhaps located near one pole, thereby producing a side-on corkscrew. Note that finding the OH or $H_2O$ to have very different spatial morphology from the other species is almost becoming common, since this phenomenon was observed in Comet 103P/Hartley 2 from the EPOXI spacecraft (A'Hearn et al. 2011) and from the ground (Knight & Schleicher 2011), in Comet Lovejoy (C/2013 R1; Opitom et al. 2015), and in Comet Lulin (C/2007 N3; Bair et al. 2018), as well as several other comets that we have observed but are as yet unpublished. In the Lulin case, two near-polar sources are required to explain the pair of opposite pointing CN corkscrew jets, but a third, near-equatorial source was required to explain some of the dust morphology and possibly the unusual OH morphology. While we will explore the OH morphology in T-G-K more thoroughly with future modeling, it is evident that the carbon-bearing species are not driven by water but some other volatile, presumably either CO or $CO_2$.

### *3.3 CN Jets Throughout the 2017 Apparition*

With the nucleus undetectable, dust images exhibiting little to no rotational variations, and the other gas species only observed at limited times, the CN morphology is our source of data for period determinations. In Figure 3 we present enhanced CN images from each of the 46 nights, all having been scaled and trimmed to 30,000 km on a side. Up to four representative images are given each night to show the motion of the CN jets with rotation, ideally spaced at intervals of approximately 2.8 hr but ranging between 1.7 and 3.6 hr; blanks are given when no useful images were obtained. Note that star trails are often evident, and sometimes appear disjointed due to the 3-5 individual images combined into each image set.

[FIGURE 3 HERE: CN from every night; up to 4 representative images per night]

An excellent example of the rotational motion during a night is evident on March 22, where a strong jet points roughly westward early on, moves towards the south-west in the second set, and essentially shuts-off in the 3$^{rd}$ and 4$^{th}$ steps. Meanwhile, a faint jet is visible towards the north-east at the start, becomes brighter and rotates towards the north by the 2$^{nd}$ step and exhibits a hook towards the east. It continues to rotate towards the northwest by the 3$^{rd}$ step, and is nearly westward by the last time step shown. Based on other nights' data, this jet also apparently nearly shuts down when it would be expected to swing through the south and towards the east. Thus both jets exhibit the same basic pattern and this diurnal cycle is what would be expected given the Sun's location towards the north-northwest (position angle, PA~343°). Another characteristic is that the apparent rate of motion of the jets clockwise around the nucleus varies with position angle, presumably due to large changes in projection effects with position angle. For instance, the dominant jet on March 27 takes significantly longer to move from north to west than it requires to rotate 90° in other directions (as seen on other nights), indicating that the jet was closest to the plane of the sky when it was in the northwest quadrant.

The major change that has taken place by the late April run is that of the overall viewing geometry, both with the comet passing the Earth, and with the Sun's PA now remaining near ~55°, as compared to values ranging from 335° up to 5° during our March run (see Table 1). Peak brightness for either jet roughly corresponds to this new direction while the jets are near minimum brightness or are not even evident in the anti-sunward direction, i.e. the southwest. Other characteristics are quite similar to those of late March, as seen in the three successive nights of April 22, 23, and 24. In each case, each jet exhibits an open spiral appearance and brightens or dims in relation to its position angle vs. the solar direction. Their apparent rotational motion also changes with PA, making a direct comparison to late March more difficult, but on average these rates are, indeed, significantly slower in April. While differing projection effects also come into play in this regard, overall the jets have somewhat less curvature in April as compared to March, consistent with a more slowly rotating nucleus.

Having only snapshots from late-May onward, we have no information regarding motion during individual nights, but other basic attributes remain – one or two jets visible, slight clockwise curvature, and generally brightest in the sunward direction. Snapshots also limit our understanding near the beginning of the apparition (from mid-February through early March), but with the context just provided by the late-apparition results, we note in particular that the curvature of the jets is much stronger in these early weeks, directly implying a shorter rotation period. More confusing, are the unusual shapes most evident on February 16 and March 9, where there is a jet apparently emitted towards the northwest, with a spiral northward and curving towards the east, i.e. consistent with the jets seen in late March, but the second jet, towards the southwest, appears to have contradictory shapes between these two nights even though the viewing geometries are similar. By combining the five nights of early snapshots, however, we tentatively conclude that this same jet remains in the southwest quadrant throughout, and that we are viewing a slightly side-on corkscrew. By late March our viewpoint moved to within the corkscrew, resulting in it also having a spiral shape. Preliminary modeling confirms that this is a viable scenario, and this subject will be examined in detail in our second paper when modeling is completed.

# 4. PERIOD DETERMINATIONS

## *4.1 Period Determinations from Pair Matching*

Our primary method for determining the rotation period of Comet T-G-K was, indeed, also the most basic – find pairs of images in which the detailed morphological structure looks essentially the same. Since measuring its rotation period was a primary goal of our observing campaign, we intentionally obtained dense temporal coverage early in our first major run in late March, both to look for jet motion within a night and to increase the probability of finding matching pairs among the nights of the run. Already discussed was our discovery of clear rotational motion within individual nights. Moreover, with about 10 hours of coverage on each of the successive nights of March 21 and 22, with 30 and 20 sets, respectively, we could easily distinguish any period ranging from 14 hr (if the last image on the 21$^{st}$ matched the first image on the 22$^{nd}$) to 34 hr (first on the 21$^{st}$ matched the last on the 22$^{nd}$). Unexpectedly, matching images occurred at nearly identical times on each night, for a period based on five pairs of 24.1 hr, and even more surprising was that this value was substantially longer than that of 19.9 hr that had just been reported by Farnham et al. (2017) based on CN imaging early in March. Note that with 10 hours of coverage per night, we not only directly ruled out periods of less than 10 hr, but even periods of up to about 50% greater than 10 hr were extremely unlikely, based on the total rotational motion during this interval. Thus, the only viable period less than 34 hr was that at 24.1 hr, with no aliases possible.

The next surprise took place only a few days later when images obtained on March 24 and 25 yielded an answer of ~25.6 hr, and subsequently pairs on the 26$^{th}$ and 27$^{th}$ gave an even longer value of ~27.1 hr. We also could, of course, find pairs throughout this entire interval, where longer baselines yielded higher precision but with the possibility that the ongoing change in viewing geometries could adversely affect the results due to associated changes in projection effects. Note that in many cases, the best match of a specific image on one night corresponded to a time intermediate to consecutive images on another night, i.e. one slightly preceding the desired match and the next one slightly following; in such cases we estimated the time at which the jets would have best matched the original image. Including snapshots obtained on March 18, 19, and 20, and on March 29 and 30, we ultimately found a total of 58 image pairs from which we could compute periods, and these are all listed in Table 2 in order of the associated mid-time. Also presented is the total interval for each pairing and the number of intervening cycles. Representative examples of these image pairs are shown in Figure 4; in some cases, where an intermediate time was computed from two consecutive images, the closest image is shown.

[PLACE TABLE 2 HERE – LIST OF ALL PAIRS]

[PLACE FIGURE 4 HERE – SAMPLE IMAGE PAIRS – 3 from March, 3 from April]

Attempts to match snapshots obtained in mid-April with any of the late March data were stymied by the rapid change in viewing geometries surrounding perigee in early April. Nightly coverage improved in later April, minimizing the time intervals between matching pairs, even though the number of hours that T-G-K was available each night was significantly less than in March. This fact, coupled with a much longer apparent period, yielded only 12 matching pairs (listed in Table 2) with yet another surprise. Not only were the derived periods longer than expected from a simple linear fit extrapolation of the late-March results, but that the rate of change of the period was clearly accelerating in late-April and very early May, as shown in Figure 5 where we plot the derived periods as a function of time.

[PLACE FIGURE 5 HERE – PERIOD VS ΔT (DATE) PLOT]

While there are no formal uncertainties using the pair matching method, we estimate that we can match most image pairs to within a few percent of the rotational phase from mid-March through early-May. During our late-March run, the typical spacing between CN images was 30-60 minutes and the rotation period was 24-27 hr, so the sampling in rotational phase was 2-4%. By late-April to early-May, the typical spacing was 1-2 hours, but the rotation period had increased to 45-50 hr, again yielding 2-4% sampling in rotational phase. However, the large number of image pairs greatly reduces the overall uncertainty during any sub-interval, as is evident in the minimal dispersion evident in Figure 5.

A difficulty throughout the apparition but particularly in the April and later time frame was distinguishing between the two jets, especially as they were roughly one-half cycle apart. In certain directions the two jets look very similar and, being about a half-cycle apart in rotational phase, we often could not determine which was which from an individual image. One solution was examining many images in context, finding those images where the jets were uniquely identifiable, and using progressions to identify them on other images. However, in other instances we needed additional clues that we discuss next.

*4.2 Period Determination from Rotational Phasing*

The need to uniquely identify each jet on all of the images to avoid period computations with a half-cycle offset from the correct value resulted in both another method for identifications along with a new technique for period determinations. The identification method was based on more subtle differences between the jets on a given image and between possible image pairs. For instance, in late March one of the jets had a distinct 'hook' when it was northward (March 22 at 6:08 UT or March 26 at 10:54 UT) while the other jet did not (March 26 at 2:48 UT or March 30 at 11:54 UT), thus confirming our identification of the two features. With the two jets looking even more similar in April, along with longer periods, distinguishing between the jets was both more difficult and more critical. Here the spacing between the jets provided the key. As an example, the last image (at 11:36 UT) shown in Figure 4 on April 20 has a dominant northern jet that looks nearly identical to the last image (at 10:45 UT) from the following night, April 21. However, the location and relative brightness of the secondary jet is quite different, pointing south-southeast on the 20$^{th}$ and east-southeast the next night. Furthermore, late on the 22$^{nd}$ (at 10:19 UT) the secondary jet matches positions with the secondary on the 21$^{st}$, but now the primary jet has rotated an additional ~30° from the prior two nights. These three nights' of images were ultimately explained, and the jets uniquely identified, using a ~42 hr period. This technique was successful throughout April, but was insufficient for dealing with many of the snapshots even later on in the apparition.

An extension of this method was to take a potential period, compute the rotational phase for each of a series of images obtained over a few days or a week's interval, and reorder the images to look for either a smooth progression of both jets throughout a rotational cycle or find discrepancies that implied the period was incorrect. This technique permitted us to determine which rotation period gave the most consistent sequence of images and provided an independent check for the rotation periods obtained by comparing single images. It proved especially useful in the April timeframe for filling in gaps between the pair-wise period solutions; these additional solutions are given in Table 3 and are distinguished in Figure 5 by orange square symbols. We also plot on Figure 5 the only other period determinations of which we are aware, including the original report of rotation of 19.9±0.15 hr by Farnham et al. (2017) in early March, the lightcurve measurement of 53±7 hr by Bodewits et al. (2018) for early May, and values based on jet rotational motion within individual nights of 30±5 hr on March 31 and of 50±10 hr on April 27 by Moulane et al. (2018); all four of these results are consistent with our own measurements, though three are not strongly constrained. Further period constraints are possible from radar from the observed bandwidth, but the specific value(s) also depend the nucleus size and shape along with the radar albedo (Howell et al. 2017; Howell et al. 2018), and final determinations have not yet been made.

[PLACE TABLE 3 HERE – PERIODS BASED ON ROTATIONAL PHASING]

*4.3 Period Constraints Based on Jet Modeling*

Under most circumstances, extending period determinations to more coarsely spaced data would be easy once an approximate period was found, and including such observations would provide a greater baseline and a more precise result. But in the case of T-G-K and its rapidly changing period, images taken near either end of the apparition are in regions of highly uncertain extrapolations. We, therefore, attempted to model the jet morphology to predict the shape and location of the jets as a function of time, first during the nearly seven weeks of our more intensive observations, and then extended throughout the apparition. At this time, our preliminary model solution, with an approximate pole orientation and source locations on the surface of the nucleus, generally works quite well during the seven-week interval, though some discrepancies remain. In particular, this preliminary solution confirms our identifications of which jet is which as discussed in the previous sub-sections. Extending the model comparisons beyond this interval, however, have thus far proven more difficult and have yielded somewhat ambiguous results, because they depend on the detailed parameters of the model that are still under investigation. Thus we defer discussing additional derived periods beyond the primary seven-week interval based on our jet model until the next paper.

*4.4 T-G-K's Rapidly Changing Period and Rotational Sequences*

As immediately evident from Figure 5, the change in our measured periods appears quite smooth but accelerates with time. Over our range of data (March 18 to May 4), we obtained an excellent fit using a $2^{nd}$ order polynomial, $0.8755 - 0.0006892 \cdot \Delta t + 0.0003262 \cdot \Delta t \cdot \Delta t$, where $\Delta t$ has units of days from March 0.0, that is very useful for interpolating values but should *not* be used for significant extrapolations, though it does nearly approach the period given by Farnham et al. (2017) in early March. In particular, we caution that this parabolic curve cannot be the overall correct solution simply because it continues to steepen with time later in the apparition, while the presumed torqueing due to outgassing driven by solar radiation *must* eventually drop off, and will also presumably fail very early in the apparition when the parabola turns back up. Rather, we assume that at both extremes the period asymptotically levels out as activity falls off when the comet is far from the Sun. With these cautions, a functional form of the fit was needed both for interpolating values of the period throughout the interval of our derived values, and to compute the correct rotational phase for each individual observation needed for making the rotational phase sequences used in Section 4.2 and in the on-going jet modeling. Rotational phases were derived by differentiating the $2^{nd}$ order polynomial, and resulting representative animation sequences from late March and late April are presented in Figure 6, having eight near-equal phase steps and with the rotational phases given at the bottom-right of each image.

[PLACE FIGURE 6 HERE – 2 ROTATIONAL SEQUENCES (LATE MAR & LATE APR)]

As the easier case to visualize and describe, we first examine the late April sequence, where one jet remains nearly in the plane of the sky for the entire rotation. Starting with this jet (labeled J1 in Figure 6), towards the ESE in the first time step, it becomes the dominant jet as it moves clockwise during the second through fourth time steps, and then disappears by the sixth step only to be faintly visible by the seventh and eighth steps. The other jet (J2), pointing towards the NNW at the first time step, rapidly fades away and then begins to reappear towards the south by the third step. It then continues clockwise, pointing east by the fifth step, and passes north between the seventh and eighth steps. The rapid move from the northwest to the south from the second to third steps is characteristic of a large projection effect as the jet nearly passes along our line of sight. When combined, these behaviors suggest that the first jet has a near-equatorial source while the second jet has a mid-to-high latitude source region, thereby yielding the asymmetric projection effect. Finally, we emphasize that all images in the late April time frame showed the same, self-consistent behavior, and the specific eight images displayed here were chosen for their near-equal spacing with rotational phase and having better S/N than other images at similar phases.

Turning back to the late March rotational sequence, several characteristics are very similar to the late April rotational cycle. At the first time step, the jet towards the east is just beginning to appear and then becomes dominant by the third step when it is towards the NNW, and eventually disappearing by the sixth step. The other jet, towards the WNW at the first time step and towards the SW at the second step, is only barely visible towards the south by the third step. Unexpectedly, it reappears towards the north by the fifth step, merging with the first jet to give a much broader fan shape. However, it remains towards the north for the next two steps and then moves towards the NW, while the first jet continues to rotate towards the south and then back towards the SE by the final time step. As in April, the second jet's behavior is readily explained with the jet moving along our line-of-sight, this time from south to north, before the swept out cone swings towards the west as it approaches the sky plane. Also, as previously noted, the progression from month-to-month is also consistent with the morphology seen in late February and early March, when the second jet appears like a side-on corkscrew towards the west, while the first jet continues to appear like a face-on spiral. Finally, we note that the late-March and late-April cycles have the same zero point for the phasing (March 0.0), simply using the parabola we fit in Section 4.4 for the periods and the resulting derived phasings, and that the only changes in bulk morphology are readily explained by the slowly evolving viewing geometry; in particular, the jet towards the east in the first time step for both sequences is indeed the same jet and, based on our preliminary modeling, apparently arises from a near-equatorial source region. This demonstrates the overall excellent fit to the rotation period evolution of our polynomial; even slight errors in the period would accumulate over time resulting in significant mismatches between morphology at comparable rotational phases.

## 5. DISCUSSION

### *5.1 Is the Apparent Change in Period Real?*

Before examining some of the consequences associated with the rapidly changing period that we deduced from our CN morphology measurements, we must address the possibility that what we observed is not due to a nucleus rapidly spinning down but instead might in some manner be caused by the nucleus actually having a non-principal axis (NPA), i.e. complex, rotational state. Preliminary work by Howell et al. (2018) has suggested an NPA state based on radar observations that apparently cannot be reconciled with our observed rapid change in rotation period unless 41P's radar albedo is unusually low. One scenario could be that the observed period is actually associated with the beat frequency between two component periods, and other scenarios might be possible. However, we are unaware of any such possibilities that would also yield the apparently simple jet morphologies that we observed throughout the apparition. Rather, any appreciable NPA motion of the nucleus should yield structures with odd and non-repeating shapes. If either component period associated with NPA were changing with time, then jet morphology would become even more complex. Moreover, we also note that the observed jets exhibited much less curvature in late April as compared to March, completely consistent with a slower rotation. Because the available data from T-G-K throughout the seven weeks presented in detail here is completely consistent with a simple, principal axis rotation state, we continue the remainder of this discussion based on this simple scenario. We acknowledge that an NPA solution may be needed in order to reconcile this with other as yet unpublished datasets, however, it is incumbent that such a model can reproduce the observed morphology reported here.

### *5.2 Torques*

As previously indicated, we were initially surprised that the value of the period continued to accelerate even into late April, given that the comet had been receding from the Sun since perihelion on April 12. However, we then realized that the effects of torques on the nucleus' rotation are better expressed in units of frequency rather than period. If we assume that the moment of inertia is unchanged, then a constant torque should yield a constant angular acceleration, yielding a straight line in frequency space. As shown in Figure 7, our observations yield a nearly linear change with time, consistent with a near-constant torque over this 7-week interval, with only a slightly shallower slope in the final two weeks. In fact, the heliocentric distance changed very little, so a near-constant torque is quite reasonable as long as there were no strong seasonal effects.

[PLACE FIGURE 7 HERE – FREQUENCY VS ΔT (DATE) PLOT]

We next turn to the torques required for the observed rapid spin-down of T-G-K's nucleus. The simple fact that the rate of change in T-G-K is about 10× greater than observed in 103P/Hartley 2 (e.g., A'Hearn et al. 2011; Knight & Schleicher 2011; Samarasinha et al. 2011; Belton et al. 2013; Knight et al. 2015) or in C/Levy (1990c, 1989 K1; Schleicher et al. 1991; Feldman et al. 1992) – these two comets having the fastest rates of change determined prior to T-G-K – strongly implies that unusual and even special circumstances were taking place. Obviously a small mass and/or a low density are required, but for outgassing to provide a large and sustained torque also requires a large lever-arm. Therefore, we consider it highly likely that the nucleus is elongated, i.e. a prolate rather than oblate spheroid. It further suggests that the locations of the source regions are not only near an end of the nucleus but facing approximately perpendicular to the long axis. These requirements could easily be met by a body having the shape of Hartley 2, but having a somewhat smaller size.

Previous studies (e.g., Samarasinha et al. 1986; Samarasinha & Mueller 2013) have shown that large changes in comet rotation periods are theoretically possible. We make a rough estimate of the amount of outgassing necessary in the case of T-G-K as follows. We assume the nucleus is elongated with a 2:1 ratio (for comparison, Hartley 2 had a 3:1 ratio; A'Hearn et al. 2011, Thomas et al. 2013) with an effective radius of 0.7 km (Tancredi et al. 2000; also consistent with the radar estimate of >0.45 km (Howell et al. 2017)) and approximate it as a cylinder having radius, $R$, of 0.5 km and length, $L$, of 2.0 km. Assuming this is in simple rotation through the short axis, it has a moment of inertia $I = ¼ \cdot M \cdot R^2 + 1/12 \cdot M \cdot L^2$, where M is the mass for which we have assumed a density of 0.5 g cm$^{-3}$. Torque, $\tau$, equals $I \cdot \alpha$ where $\alpha$ is the angular acceleration. We take the conditions in late March, when the period was ~24 hr and the rate of change in the period was ~0.5 hr day$^{-1}$, to find a net torque of ~5×10$^9$ g m$^2$ s$^{-2}$ radians needed to produce the observed change in rotation period. The torque due to outgassing can be expressed as $\tau_{gas} = F_{gas} \cdot d$ where $d$ is the effective distance from the center, i.e., approximately ½ L, and $F_{gas}$ is the force produced by outgassing. $F_{gas} = Q_{gas} \cdot M_{gas} \cdot v_{gas}$ where $Q_{gas}$ is the production rate (in molecules per second), $M_{gas}$ is the mass of gas molecules, and $v_{gas}$ is the velocity of the gas as it leaves the nucleus. For this order of magnitude estimate, we assume the outgassing consists entirely of $H_2O$ ($M_{gas}$ = 18 g mole$^{-1}$) and conservatively estimate that it leaves the nucleus at 0.1 km s$^{-1}$ (significantly below the canonical 1 km$^{-1}$ outflow velocity seen beyond the inner acceleration region). Setting this equal to the torque inferred from the change in velocity requires $Q_{gas}$ ~1.4×10$^{27}$ mol$^{-1}$. Moulane et al. (2018) reported $Q_{gas}$ ~ 2×10$^{27}$ near perihelion in 2017, suggesting that the outgassing could readily account for the observed rotation period change if most of the outgassing is asymmetric and contributes to the torqueing. Less conservative assumptions (e.g., higher gas velocity, dust entrained with gas, heavier gas molecules) might change this number by a factor of a few, making it easier to produce the required torque, and do not alter the principal result: the observed change in rotation period of T-G-K can be readily explained by known and likely properties of the nucleus and associated outgassing.

### 5.3 The Future for T-G-K

While we see no evidence for NPA within the 6 weeks discussed here, we note that predicting the future rotational behavior is quite difficult. Having two source regions and, based on our preliminary modeling, with one located near the equator but the other at a mid-or-high latitude, the nucleus would be expected to slow sufficiently that the non-equatorial forces might eventually produce a tumbling of the nucleus. Otherwise, in the much simpler scenario, the nucleus would simply continue to slow until it stops and begins rotating in the opposite direction due to the continued outgassing from the source regions. A simple extrapolation of the linear fit shown in Figure 7 suggests that the rotational frequency could have hit zero by +56 days from perihelion or about June 8. However, the data by late April imply a slight flattening of the curve, presumably due to the outgassing beginning to drop off resulting in a decrease in the torque. We conclude that while T-G-K might have stopped rotating late in the 2017 apparition, a similar amount of torqueing during the next approach to the Sun will easily reach a stopped and reversal of motion or a tumbling stage early in the 2022 apparition. We note that similar calculations were also made by Bodewits et al. (2018). Unfortunately, the next perihelion passage of T-G-K in late 2022 is extremely poorly placed for viewing from Earth. One may need to wait until the much better apparition of early 2028 to resolve T-G-K's fate.

### *5.4 The Past for T-G-K*

Any extrapolation of T-G-K's rotation into the past implies the comet had an ever-faster rotation rate and a corresponding shortening of its period. Again, we expect the greatest rate of change would occur near perihelion each orbit and the rate of change of the period would decrease as more and more force would be required to alter the spin rate at progressively shorter periods. Making simple but reasonable assumptions based on the 2017 apparition, we estimate that the nucleus would have been near or at the break-up rotational velocity 3-4 orbits ago. Since this coincides with the last major outburst (>7 magnitudes) in 2001 for T-G-K, we speculate that the outburst might have been caused by a partial breakup at that time. Moreover, if the entire cycle (spin down and then spin up in the opposite direction) repeats, this could further explain the pair of major (>8 magnitudes) outbursts in 1973. A variety of mechanisms for large cometary outbursts have been suggested over the past few decades, but this may be the first evidence of a specific method tied to a specific object.

## 6. SUMMARY AND FUTURE PLANS

### *6.1 Summary*

We have presented results from our extensive imaging campaign of Comet 41P/Tuttle-Giacobini-Kresák during the first half of 2017 – its best apparition since its first discovery in 1858. With the comet remaining at less than 0.15 AU of Earth for three weeks and within 0.20 AU over a two month interval, the circumstances allowed us to study its coma morphology in search of possible jets, whose appearance and motion as a function of time could yield fundamental properties of the nucleus, including its rotation period. This was indeed the case, though our discovery that T-G-K's period was rapidly changing also required a change in observing plans, requiring many more nights of data acquisition. Overall, imaging was obtained on a total of 47 nights between February 16 and July 2, using Lowell Observatory's 4.3-m Discovery Channel Telescope, the Hall 1.1-m telescope, and the robotic 0.8-m telescope, with 33 nights during the intensive interval from March 17 to May

4. Narrowband CN images were obtained on 46 of the 47 nights, and all narrowband CN images exhibit either one or two gas jets. One of the jets always appears as a simple partial spiral (turning on and off with the diurnal cycle) with clockwise rotation, while the second jet exhibits the same behavior during early April, but has a less complete spiral appearance earlier and later, and evolves from and back to a corkscrew very early and very late in the apparition. All data are consistent with a slow, steady evolution of the jet morphology from mid-February to early May and possibly longer, presumably due to viewing geometry changes coupled with seasonal changes. In particular, the detailed evolution of the jets' CN morphology suggests that the first source region is located near the equator (which would yield open spirals unless the Earth is near the plane of the equator), while the second jet is emitted from a mid-to-high latitude source region producing a jet that traces out a cone that we observe from differing viewing locations as the comet passed the Earth. Other gas species – OH, $C_3$, and $C_2$ – were observed on a much more limited basis, with the pure carbon species generally looking similar to CN, while the OH morphology generally looked quite different and exhibited very little evidence of rotational motion. Dust morphology differed from all of the gas species, showing a strong dust tail but also a much fainter, near-linear sunward feature through much of the apparition.

The clear repetitive nature of the motion of the two jets made it relatively easy to determine the apparent (rather than sidereal) rotation period. Most surprising was the result that the initial period we determined of 24.1 hr (for March 21&22) was increasing by about 0.5 hr each day, reaching about 27 hr near the end of the month. Images from April 15 to May 4 yield an accelerating change in period, passing 48 hr on about April 28. Using multiple techniques, we ultimately made 78 period determinations during the seven-week interval, and the more than a factor of two increase of the period is a rate of change about 10× greater than observed for any other comet. Since the simply and smoothly evolving jet morphology appears very difficult if not impossible to have with non-principal axis rotation, we must conclude that rotation of T-G-K's nucleus has spun down by our measured rate. While this is by far the fastest rate of change ever measured for a comet nucleus, the torque required is readily within what can exist given other known or likely properties of the nucleus. In particular, we have shown that a small, elongated nucleus with source regions placed near the end of the long-axis and oriented perpendicular to the rotation axis can yield the necessary torque based on the measured gas production.

Using the measured rate of change of the angular rotation rate, we estimate that the nucleus could have stopped rotating and/or begun to tumble (due to one source being located away from the equatorial zone) as soon as early June in 2018, only two months following perihelion, and would certainly reach this stage by early in the next, 2022, apparition. Working backwards in time, T-G-K would have been rotating near its rotational break-up velocity 3-4 orbits earlier, suggesting that its extreme (>7 mag) outburst observed in 2001 might have been caused by a partial fragmentation at that time. Successive spin-downs and spin-ups could also have contributed to the earlier pair of extreme (>8 mag) outbursts observed in 1973. Clearly, Comet 41P/Tuttle-Giacobini-Kresák deserves detailed observing campaigns at each of the next several apparitions.

### *6.2 Future Plans*

While awaiting future observing opportunities, in the nearer term we will continue work on our modeling of the CN jet morphology to further pin down the orientation of T-G-K's rotation axis and the specific source location associated with each jet. With a more definitive nucleus solution, we will then be able to use the snapshot observations obtained earlier and later in the apparition to determine the associated rotational phases for each image and, in turn, derive the rotational periods required to reach each phase, thereby extending the period vs time plot in both directions. Our next paper will also incorporate the various results we obtained regarding T-G-K's production rates and composition and to what extent these vary with seasons based on our nucleus model solution.

### ACKNOWLEDGEMENTS

We thank B. Skiff, L. Bright, and C. Trujillo for assisting with or making observations available. We thank T. Farnham for making available his period determination prior to the start of our own primary observing run, and thank B. Mueller, N. Samarasinha, and E. Howell for useful conversations regarding radar results and possible NPA motion. These results made use of the Discovery Channel Telescope at Lowell Observatory. Lowell is a private, non-profit institution dedicated to astrophysical research and public appreciation of astronomy and operates the DCT in partnership with Boston University, the University of Maryland, the University of Toledo, Northern

Arizona University and Yale University. The Large Monolithic Imager was built by Lowell Observatory using funds provided by the National Science Foundation (AST-1005313). This research was supported by NASA Planetary Astronomy grants NNX14AG81G and 80NSSC18K0856, and the Marcus Cometary Research Fund.

FIGURE CAPTIONS

**Figure 1.** Representative images of 41P/Tuttle-Giacobini-Kresak in the CN (right column) and R-band (left column) filters throughout the 2017 apparition, following enhancement as described in Section 2.2. Brighter regions are shown as yellow or white, respectively. Date and mid-times are given on each image and the direction of the Sun is indicated by an arrow; each images is 30,000 km on a side and star trails are often visible. One or two CN features are evident throughout the apparition, often appearing as clockwise rotating partial spirals, and the near-disappearance of the jets in the anti-sunward hemisphere strongly suggests that the source regions mostly shut-off at night. As expected, the R-band images show a strong dust tail, but also sometimes exhibit a fainter near-linear feature somewhat offset from the sunward direction. Overall, the long-term trends appear to be simply caused by changing viewing geometries.

**Figure 2.** Sample sets of gas images (OH, CN, $C_3$, and $C_2$; by columns) at two representative rotational phases during our late-March run and a single rotational phase during our late-April run (by rows). The brightest features are shown in yellow. Due to strong underlying continuum in the $C_3$ and $C_2$ filters, only photometric nights were used and the continuum was subtracted from these filters prior to image enhancement. The scale is again 30,000 km on a side, with the direction to the Sun indicated with an arrow. As usual, based on many other comets, the $C_3$ is nearly identical to the CN except that it does not extend as far from the nucleus because of its much shorter lifetime, while the $C_2$ features appear broader than those of CN due to having multiple parents and/or grandparents. However, the OH is extremely different, both remaining within a quadrant each night and not exhibiting significant rotation during a night, but changing its bulk orientation somewhat over several nights.

**Figure 3.** Representative CN images from every night of observations throughout T-G-K's apparition. Up to four images are presented each night to show the rotational motion. Note the slow progression of the peak brightness of the jets as they follow the Sun's position angle (arrows) implying that peak brightness is always near local noon, and other bulk changes with our viewing geometry, but that no abrupt changes in morphology are evident. Scaling is the same as previous figures.

**Figure 4.** Sample CN image pairs used to derive the apparent period. In each case shown, the matching images must be a whole number of rotational cycles apart. Scaling is the same as previous figures.

**Figure 5.** Derived periods as a function of time. The derived periods listed in Table 2 based on matching CN image pairs are shown as filled blue circles, while periods based on smooth phase cycles listed in Table 3 are given as filled orange squares. We estimate our uncertainties to be less than a few percent of the period, consist with the small amount of scatter visible. Other published period determinations are shown with open symbols (see text for the various techniques employed). The dashed parabola is the fit to our seven weeks of data, while the extrapolated curve (shown as dotted) simply continues the curve but must break down when the comet is further from the Sun and torques are reduced. Perihelion occurred on April 12, and is marked by the vertical dashed line.

**Figure 6.** A sample rotational sequence is shown for each of our two primary runs in late March and late April. Each sequence starts with the upper-left image and moves clockwise; the rotational phases are given in the lower-right corner of each frame. In each case, the jet approximately towards the east (left) in the first time step **(labeled J1)**, completes a simple, near face-on spiral through the sequence, while the other jet **(J2)** exhibits significantly larger projection effects, presumably due to having a source region closer to the rotation pole resulting in a narrower cone being traced out and, during both time frames, cutting along our line of sight with one side of the swept-out cone.

**Figure 7.** Derived frequencies as a function of time. The data points and symbols are the same as from Figure 5, except the three points having large uncertainty and thus are non-constraining are excluded here. For a constant torque, one would expect a linear progression of the frequency with time, and can more readily estimate when the comet could have stopped rotating. Based on our fit, this could have occurred only 56 days past perihelion or as early as June 8; however, there is some evidence that the slope was somewhat less steep in the final two weeks, implying a later critical date.

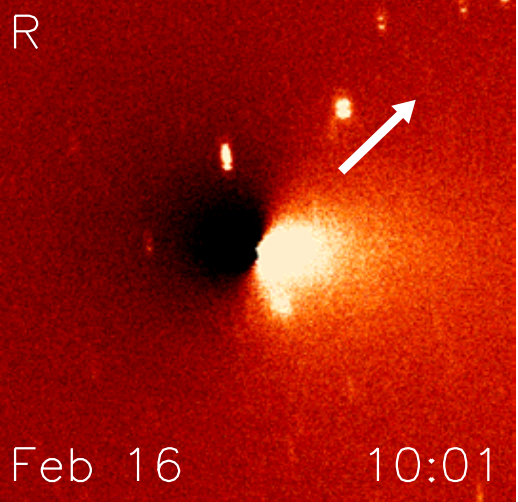
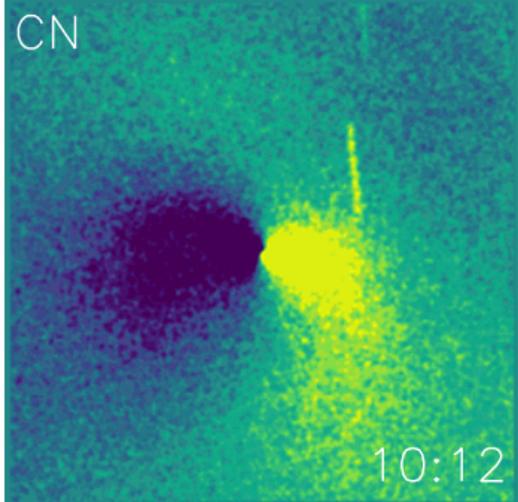
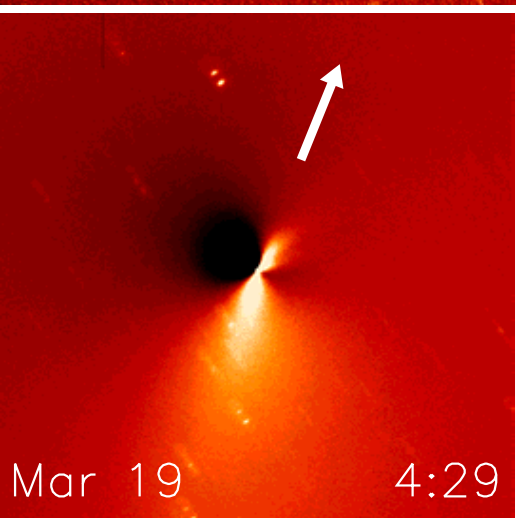
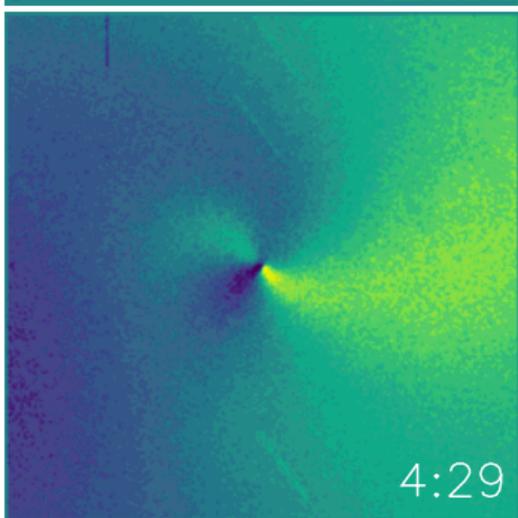
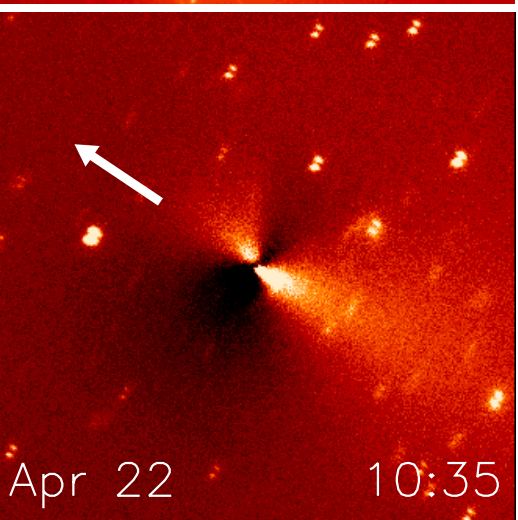
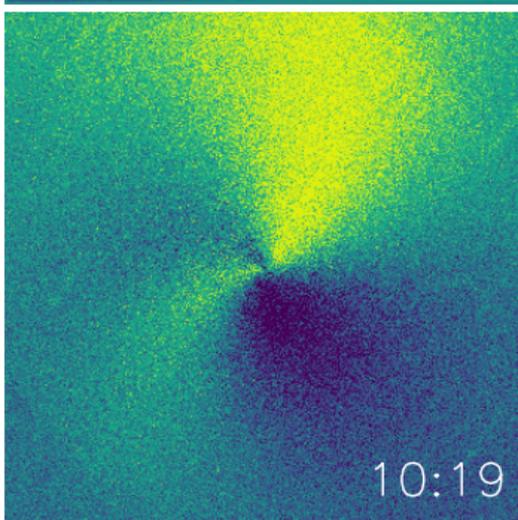
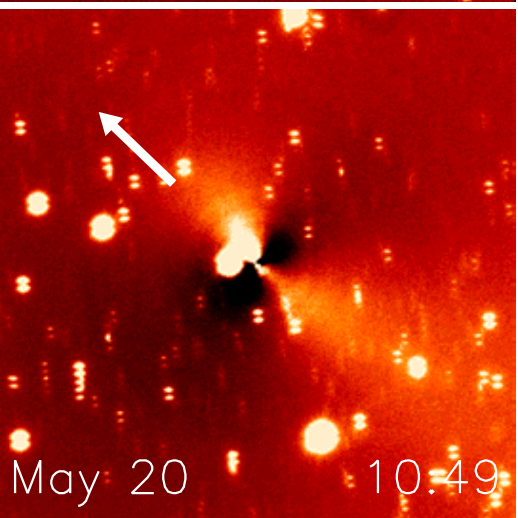
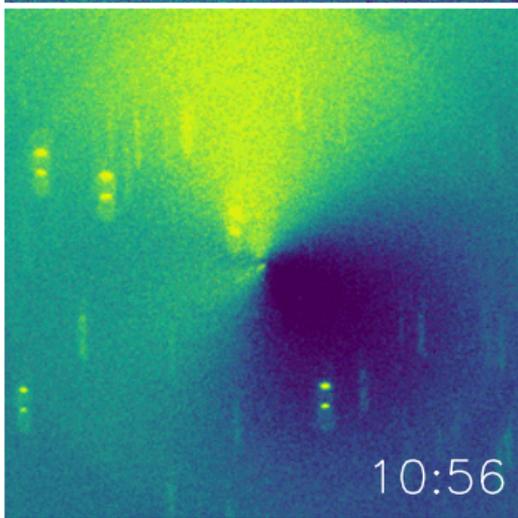
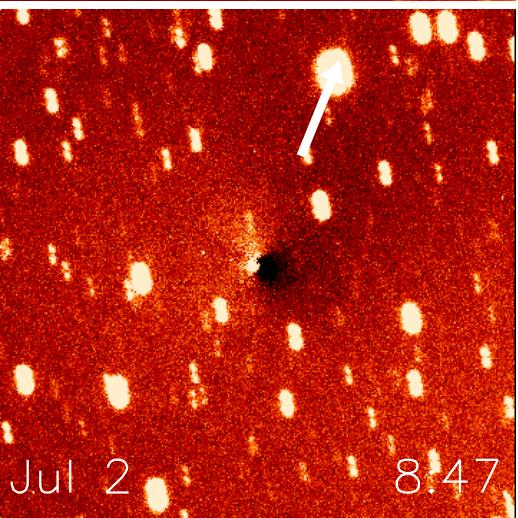
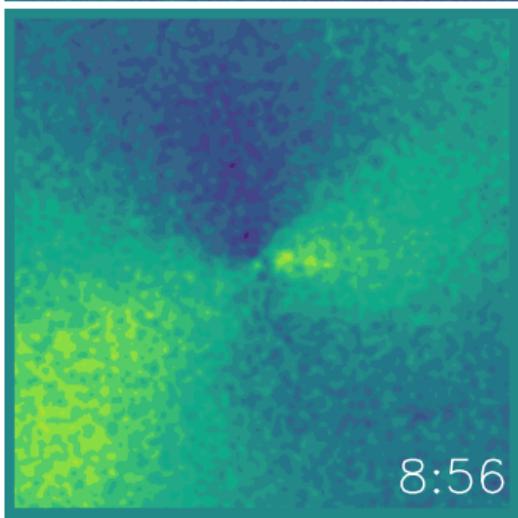

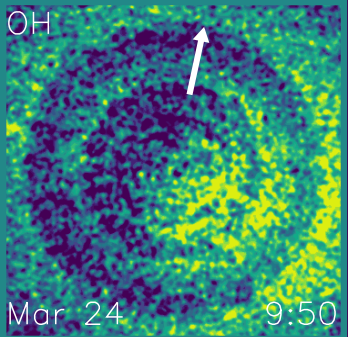 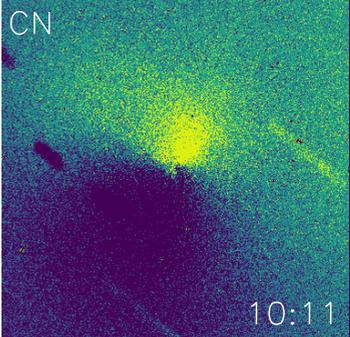 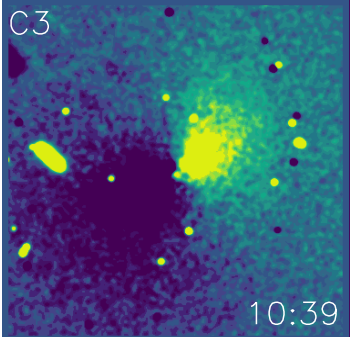 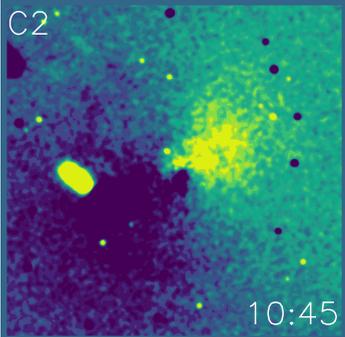
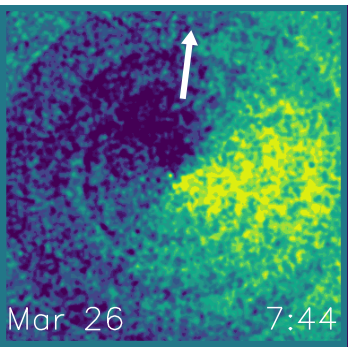 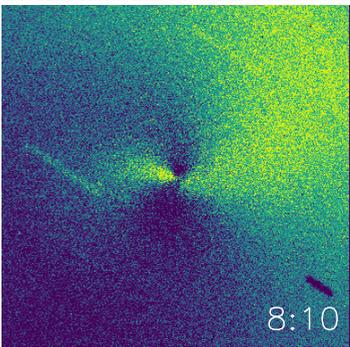 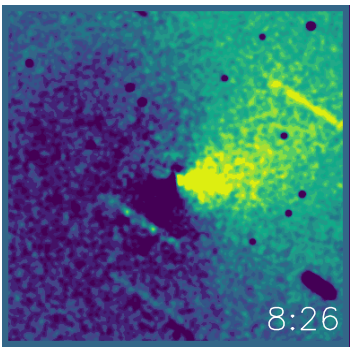 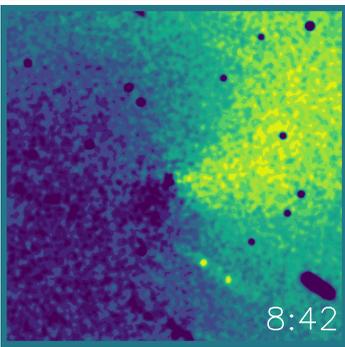
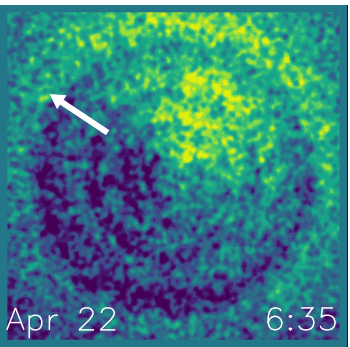 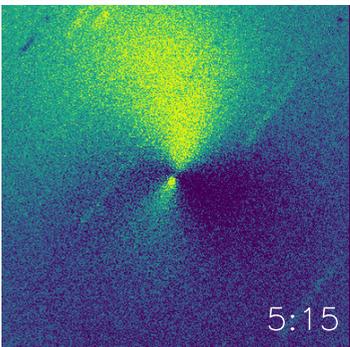 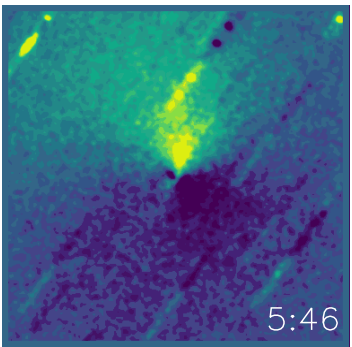 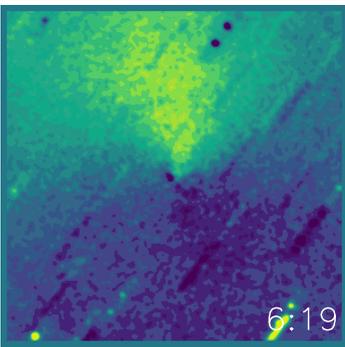

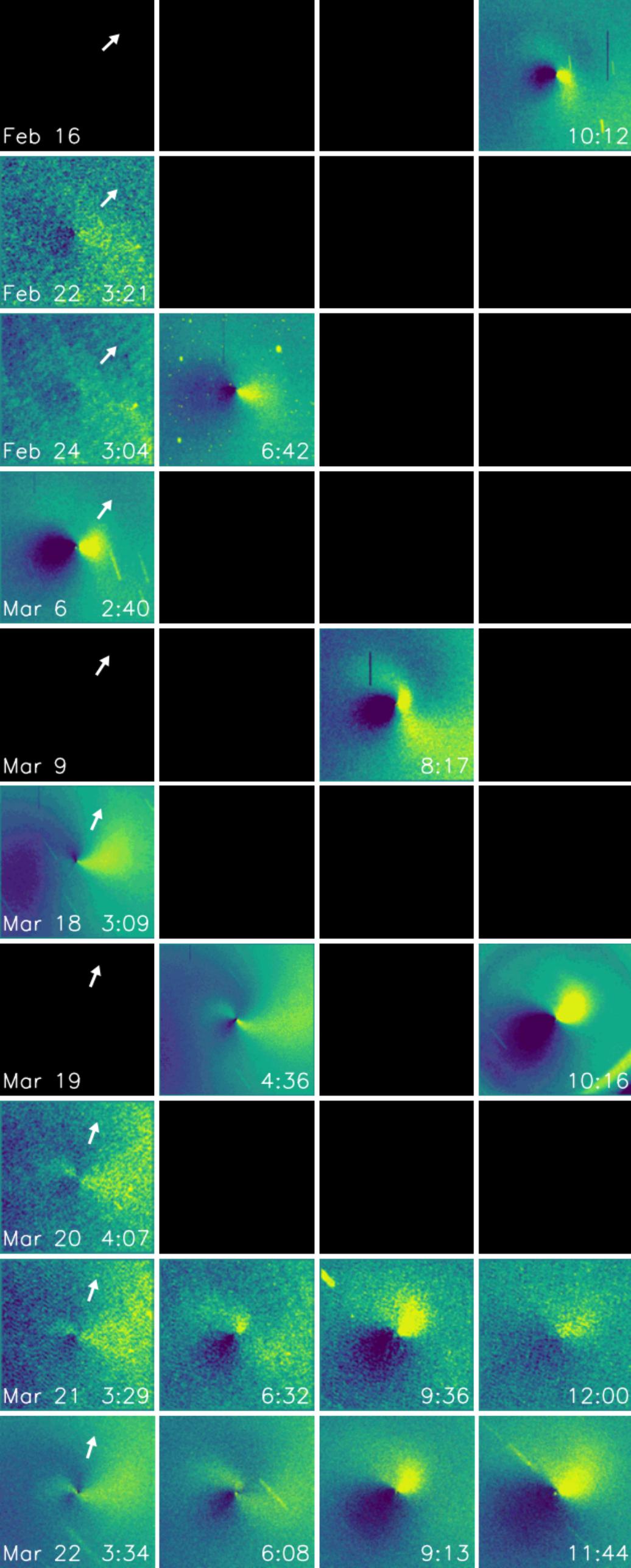

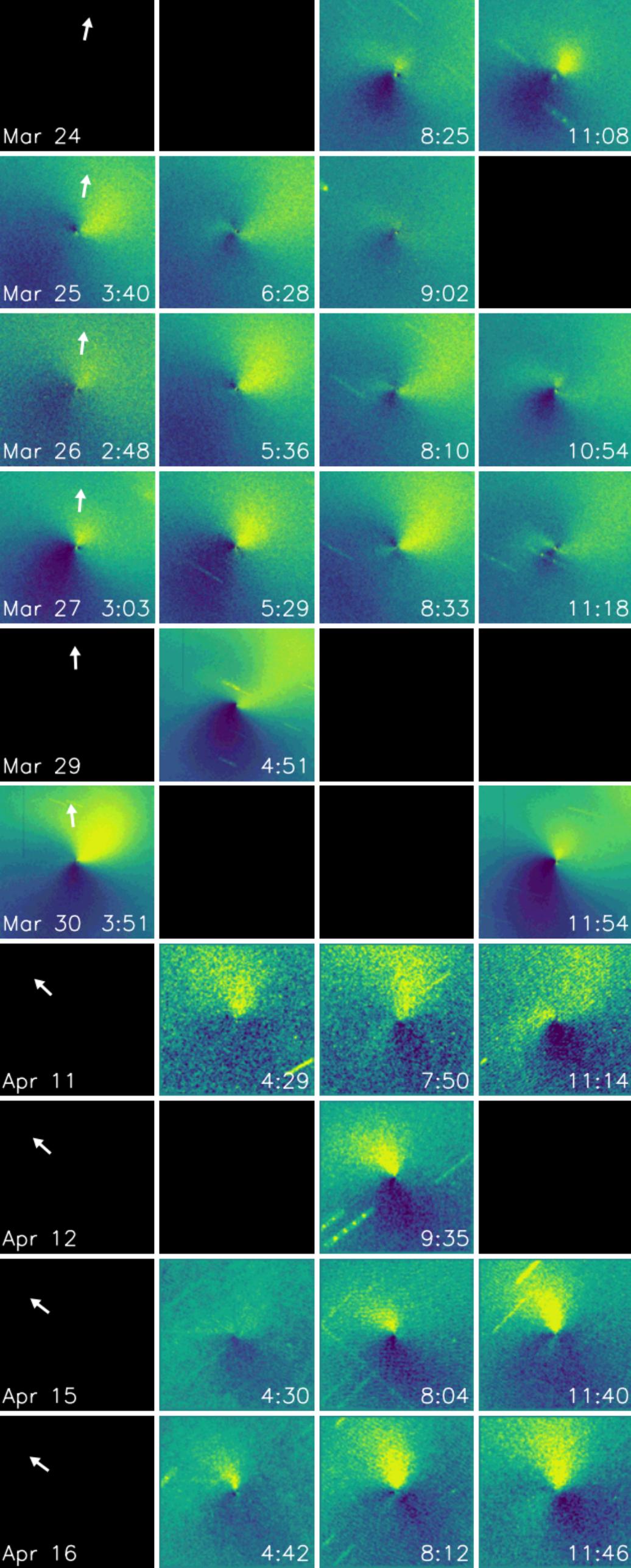

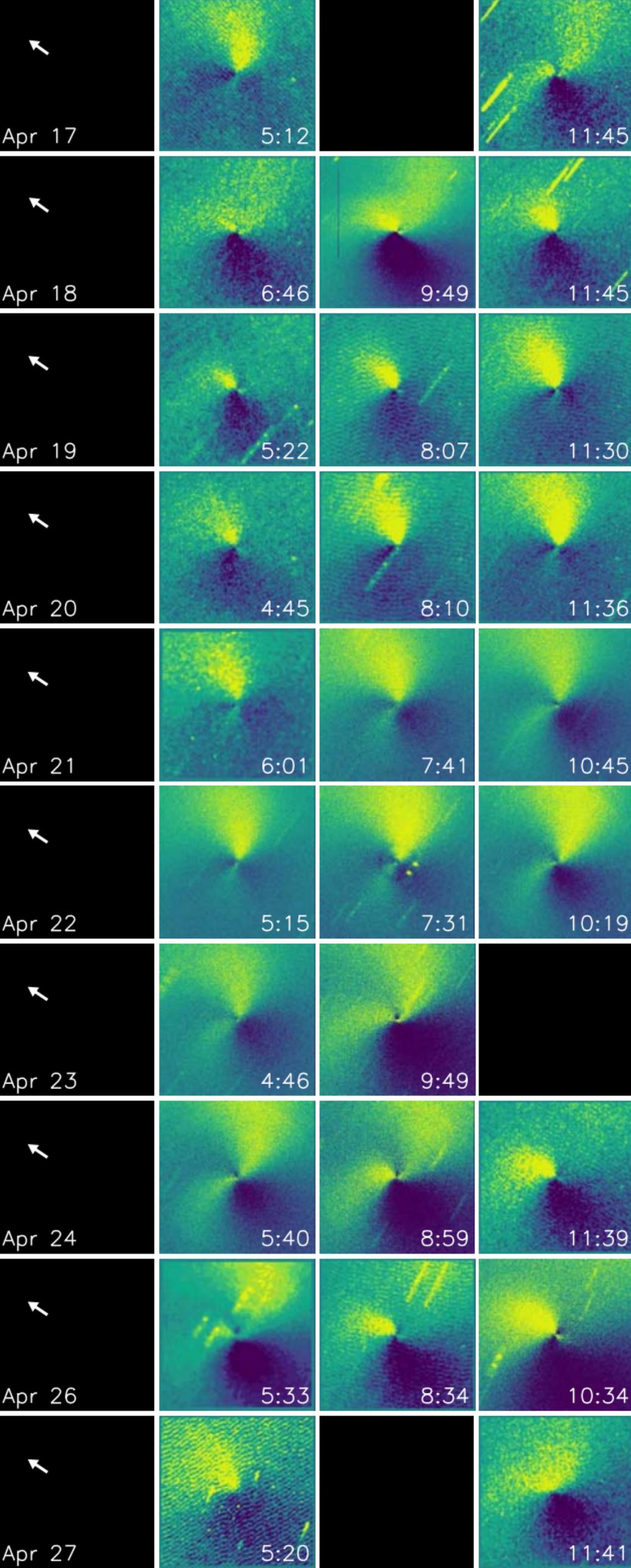

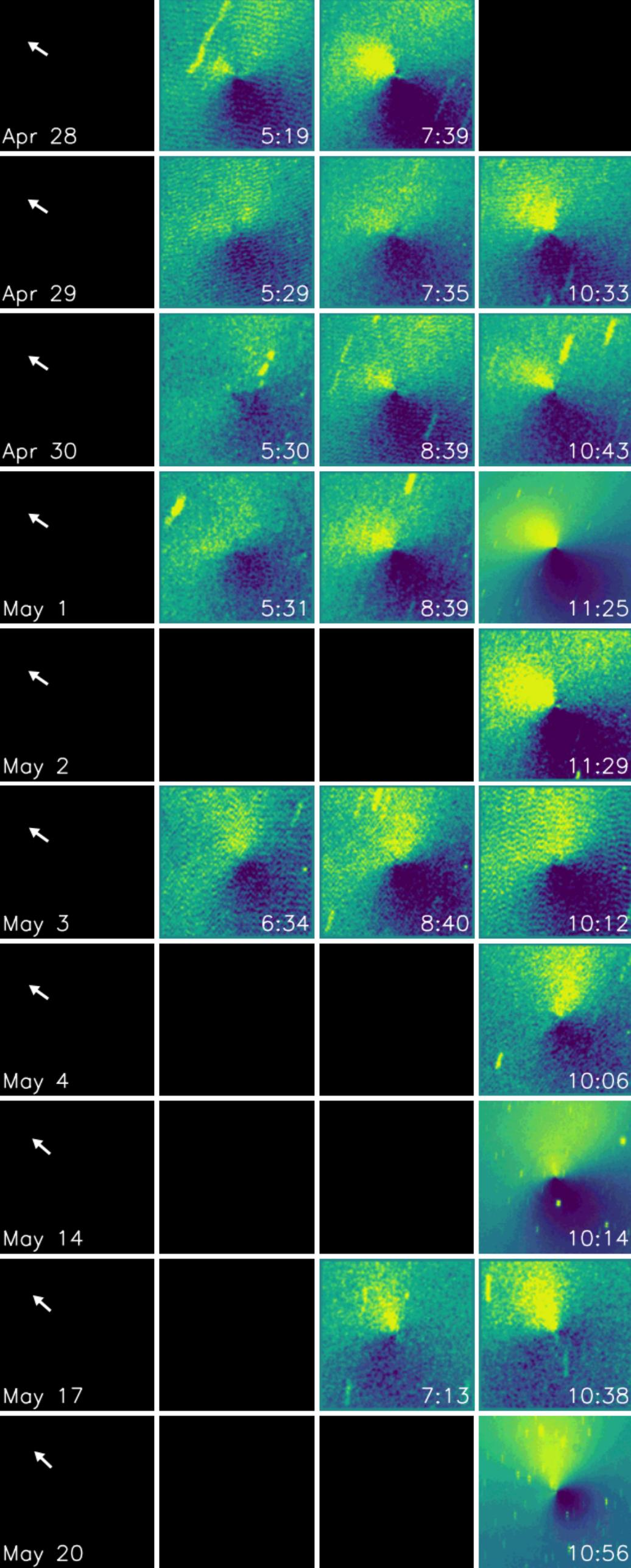

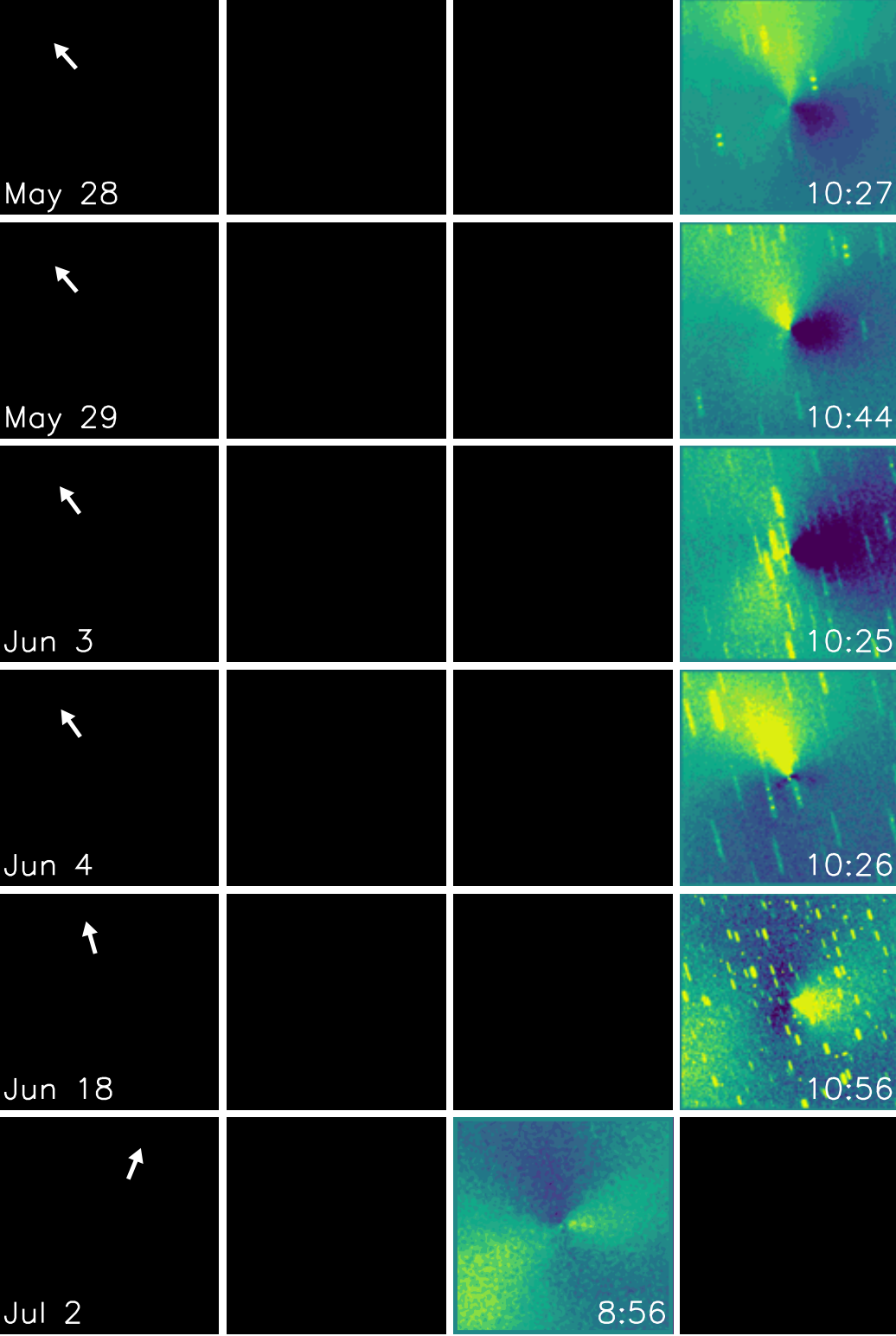

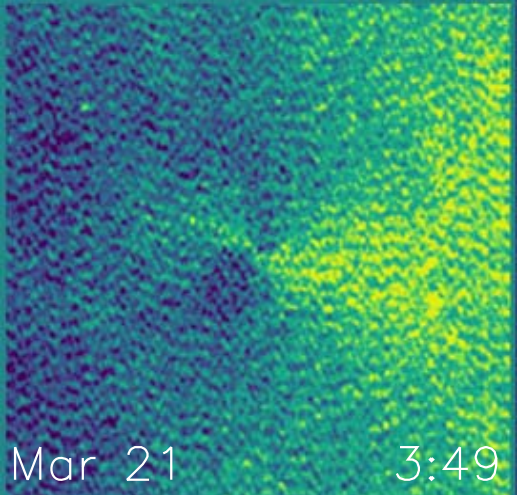
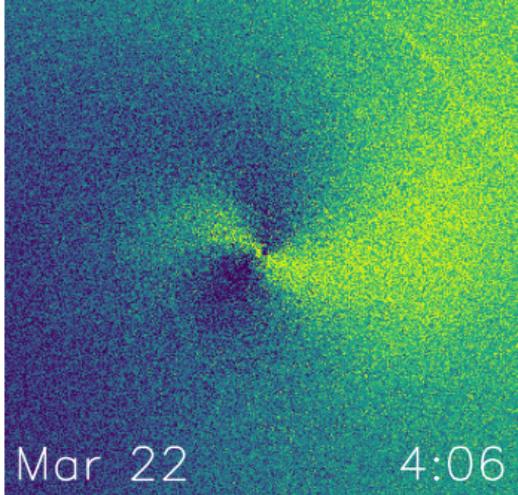
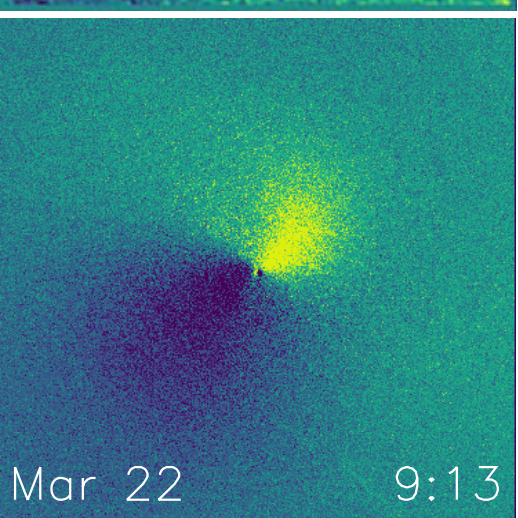
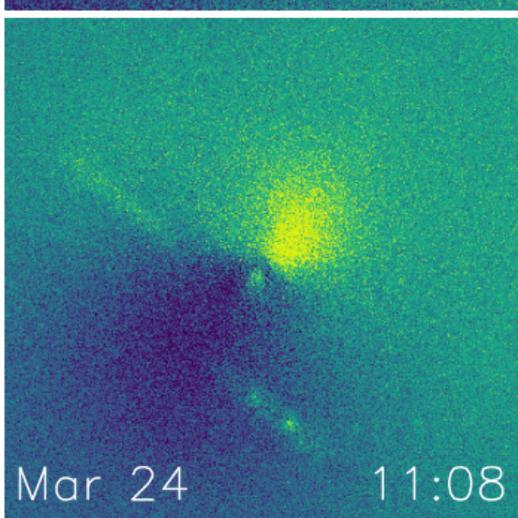
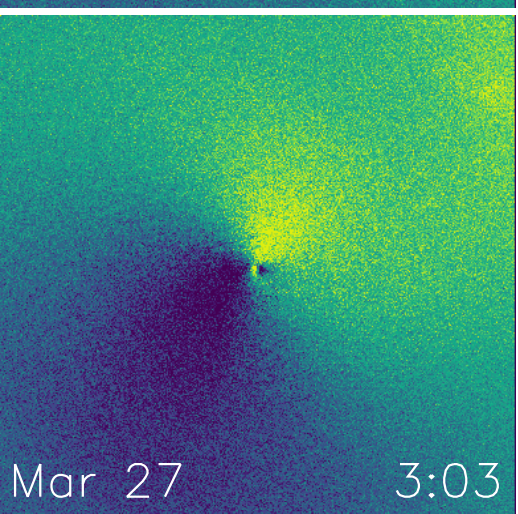
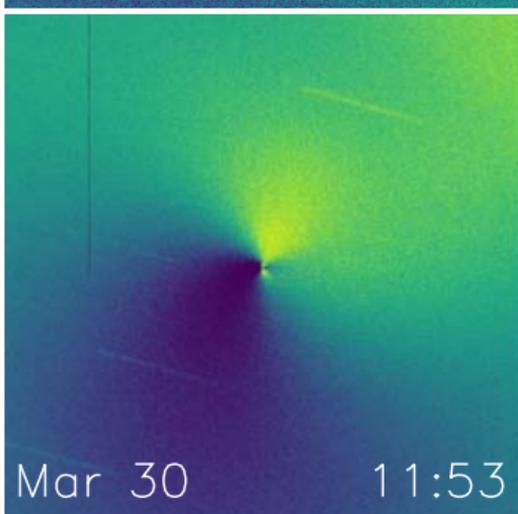
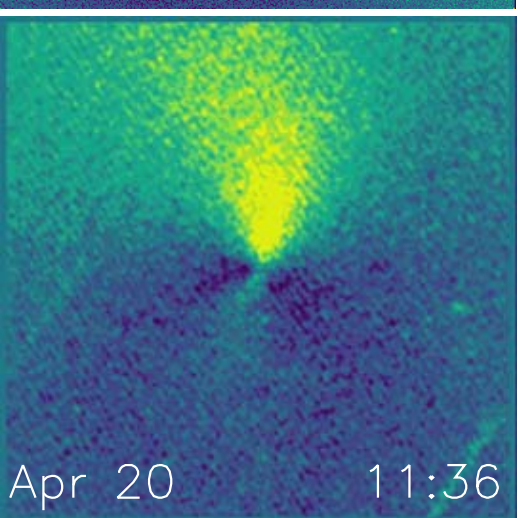
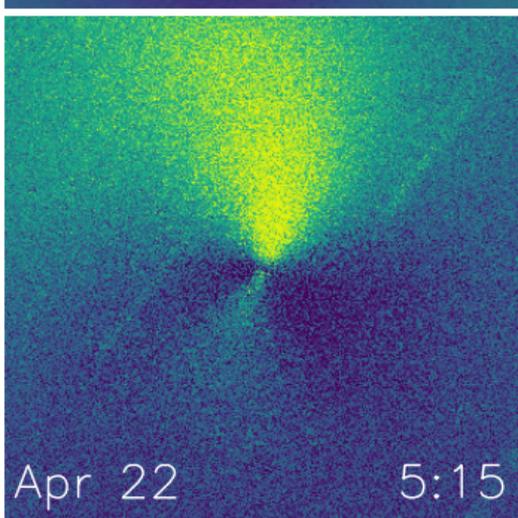
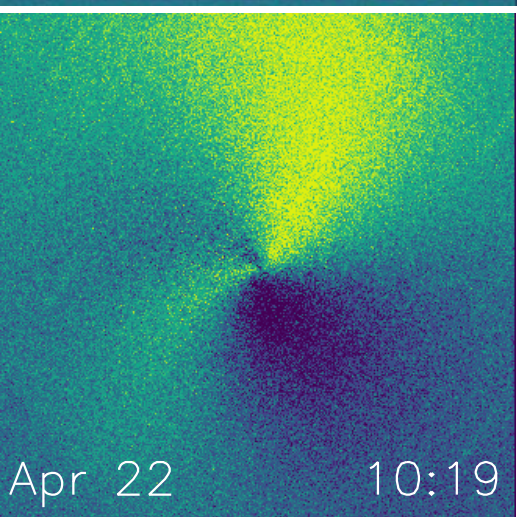
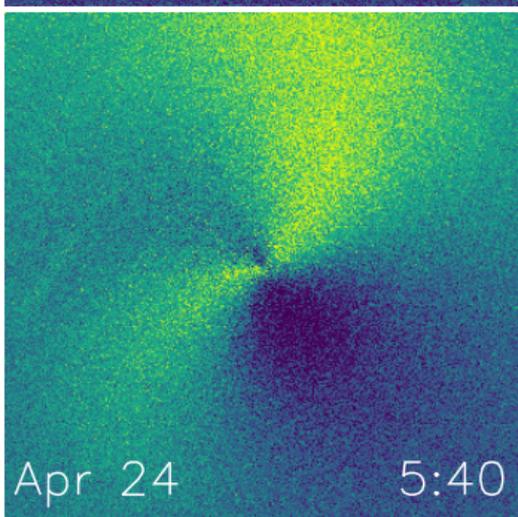
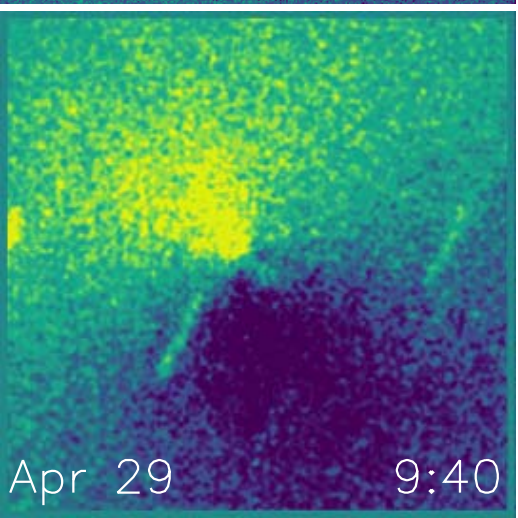
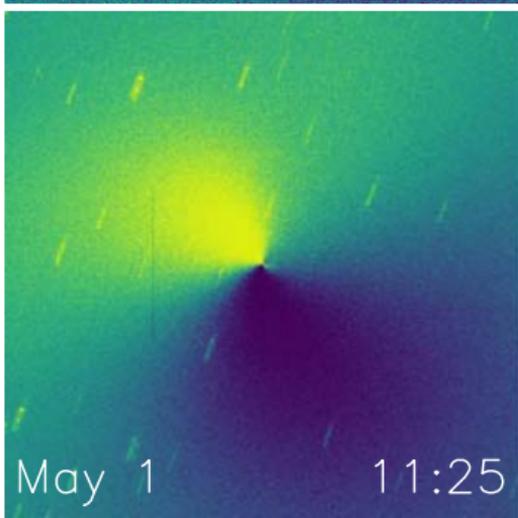

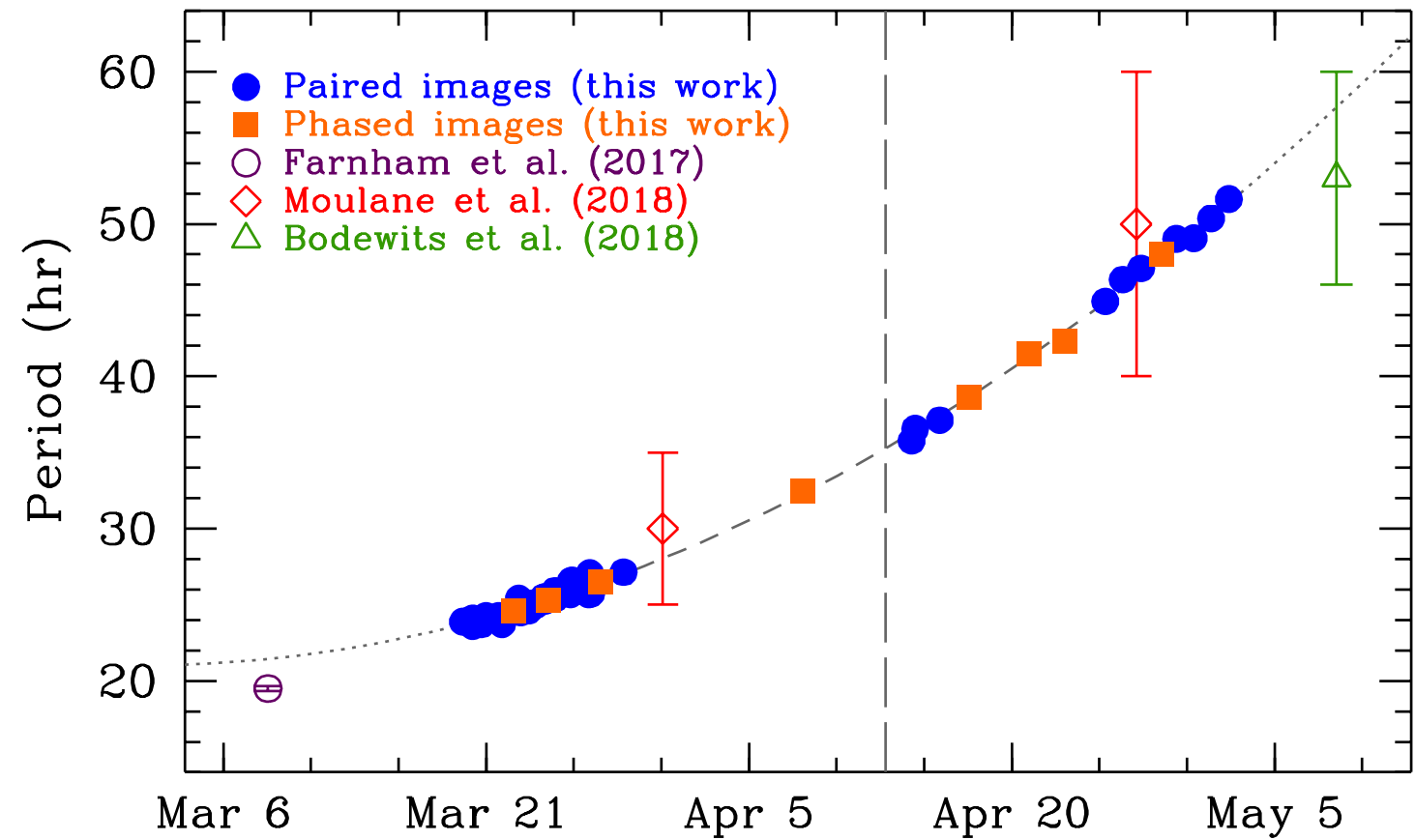

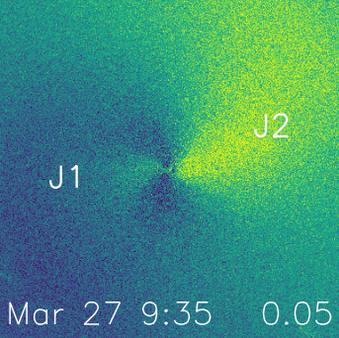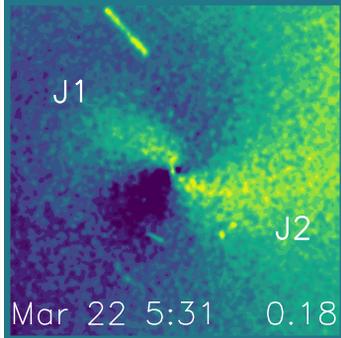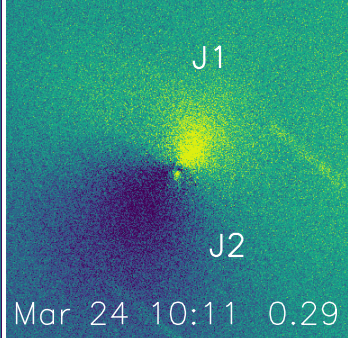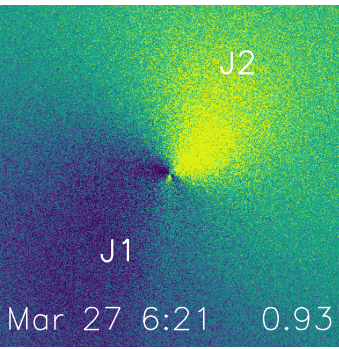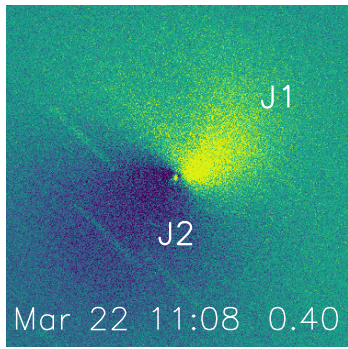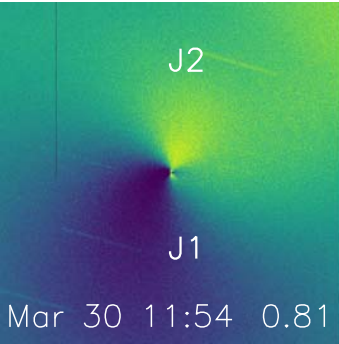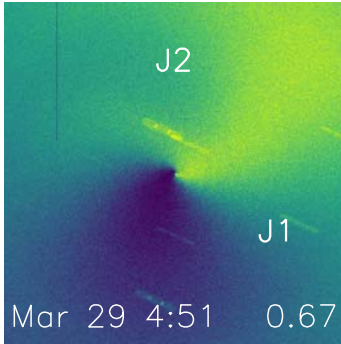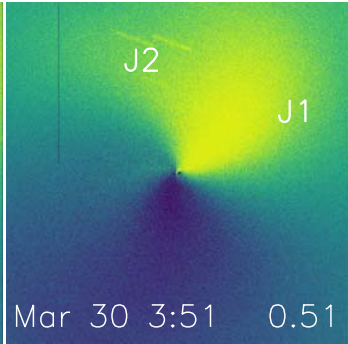

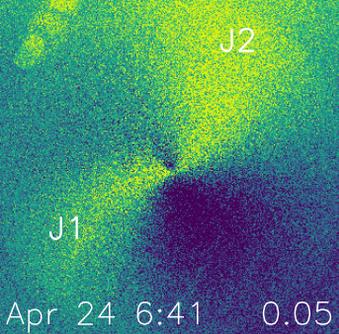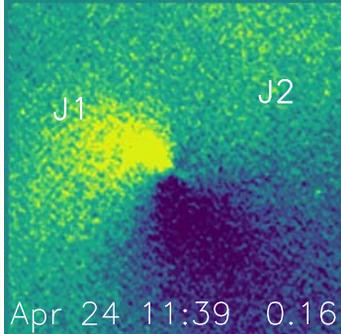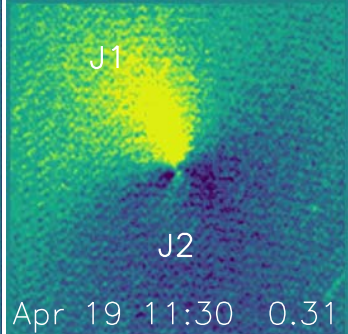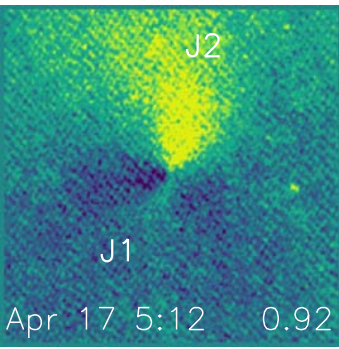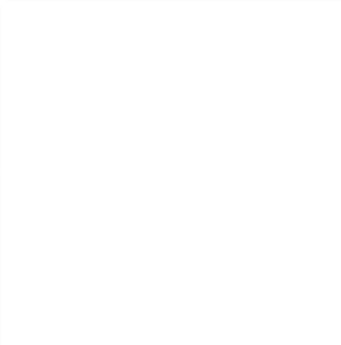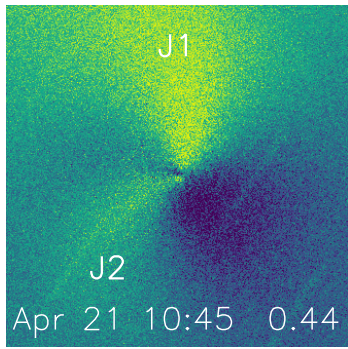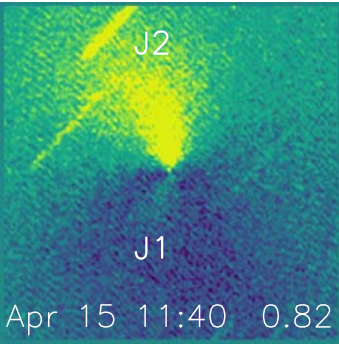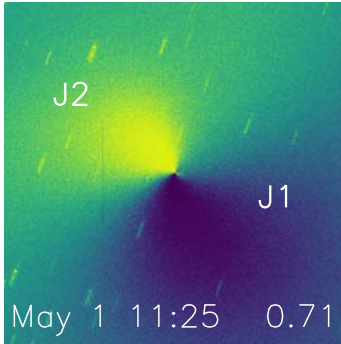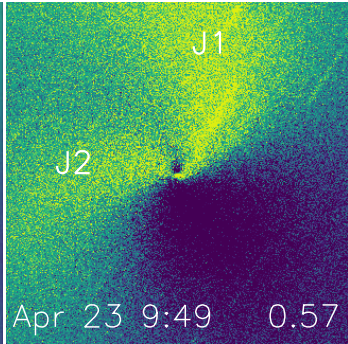

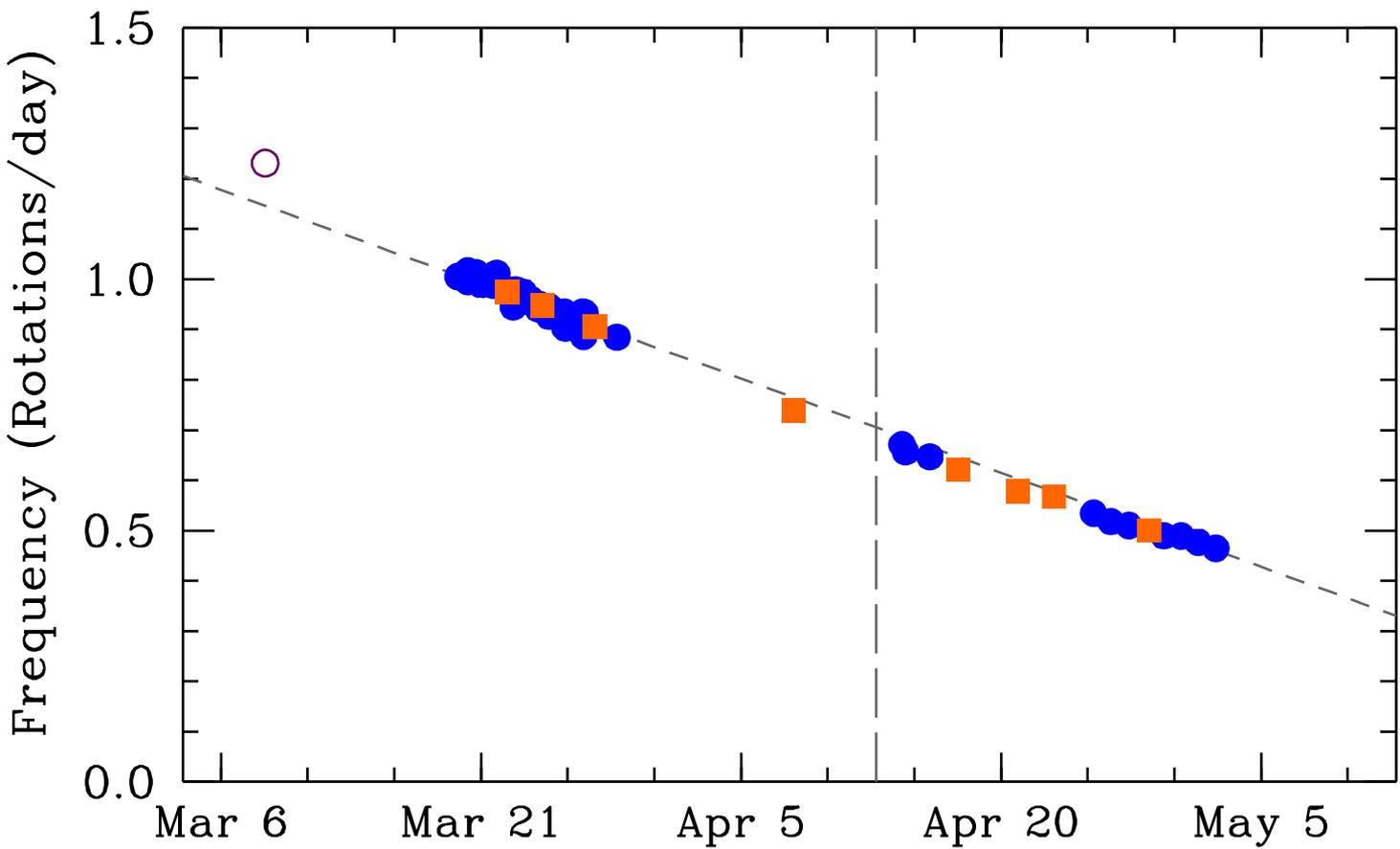

Table 1. CCD observing circumstances for Comet 41P/Tuttle-Giacobini-Kresák in 2017.[a]

| Date | UT Range | Tel.[b] | $\Delta T$[c] (day) | $r_H$[d] (AU) | $\Delta$[e] (AU) | $\alpha$[f] (°) | Sun P.A.[g] (°) | Filters | Mode | Weather |
|---|---|---|---|---|---|---|---|---|---|---|
| Feb 16 | 9:59–10:25 | DCT | −55.327 | 1.278 | 0.292 | 6.3 | 313.8 | R, CN | snapshot | photometric |
| Feb 22 | 5:47– 6:17 | 31in | −49.501 | 1.237 | 0.255 | 12.6 | 317.0 | R, CN | snapshot | clouds |
| Feb 24 | 6:39– 6:45 | DCT | −47.473 | 1.223 | 0.244 | 14.9 | 318.1 | R, CN | snapshot | intermittent clouds |
| Feb 24 | 5:09– 6:08 | 31in | −47.517 | 1.223 | 0.244 | 15.0 | 318.1 | R, CN | snapshot | photometric |
| Mar 6 | 2:21– 2:49 | DCT | −37.645 | 1.162 | 0.196 | 27.5 | 324.3 | r', CN | snapshot | cirrus |
| Mar 9 | 8:01– 8:26 | DCT | −34.410 | 1.144 | 0.184 | 32.0 | 326.8 | r', CN | snapshot | cirrus |
| Mar 17 | 5:07– 5:10 | DCT | −26.538 | 1.105 | 0.160 | 43.2 | 334.9 | r' | snapshot | clouds |
| Mar 18 | 3:00– 3:14 | DCT | −25.622 | 1.102 | 0.158 | 44.5 | 336.2 | r', CN | snapshot | clouds |
| Mar 19 | 4:27–10:21 | DCT | −24.444 | 1.097 | 0.155 | 46.2 | 337.9 | r', CN | monitoring | photometric |
| Mar 20 | 8:22– 8:31 | 31in | −23.401 | 1.092 | 0.153 | 47.7 | 339.6 | R, CN | snapshot | photometric |
| Mar 21 | 2:35–12:51 | 31in | −22.431 | 1.089 | 0.151 | 49.1 | 341.3 | R, CN | imaging | clouds |
| Mar 22 | 4:02–12:31 | 31in | −21.408 | 1.085 | 0.150 | 50.4 | 343.2 | R, CN | monitoring | clouds |
| Mar 22 | 2:42–12:29 | 42in | −21.436 | 1.085 | 0.150 | 50.5 | 343.3 | R, CN, OH, C3 | imaging | clouds |
| Mar 24 | 8:17–11:54 | 42in | −19.332 | 1.078 | 0.147 | 53.3 | 347.8 | R, CN, OH, C3, C2, BC | imaging | photometric |
| Mar 25 | 2:44–11:25 | 42in | −18.458 | 1.075 | 0.146 | 54.5 | 350.0 | R, CN, OH, C3, C2, BC | imaging | photometric |
| Mar 26 | 2:36–11:40 | 42in | −17.455 | 1.072 | 0.145 | 55.8 | 352.7 | R, CN, OH, C3, C2, BC | imaging | photometric/cirrus |
| Mar 27 | 2:46–11:57 | 42in | −16.446 | 1.069 | 0.144 | 57.0 | 355.6 | R, CN, OH, C3, C2, BC | imaging | cirrus |
| Mar 29 | 4:43– 4:54 | DCT | −14.552 | 1.064 | 0.143 | 59.3 | 1.7 | r', CN | snapshot | photometric |
| Mar 30 | 3:38–12:01 | DCT | −13.426 | 1.061 | 0.142 | 60.5 | 5.7 | r', CN | monitoring | cirrus |
| Apr 11 | 4:18–11:22 | 31in | −1.426 | 1.045 | 0.149 | 69.3 | 45.5 | R, CN | monitoring | cirrus |
| Apr 12 | 9:19– 9:48 | 31in | −0.354 | 1.045 | 0.151 | 69.6 | 47.5 | R, CN | snapshot | photometric |
| Apr 15 | 4:13–11:53 | 31in | +2.583 | 1.046 | 0.156 | 70.1 | 51.7 | R, CN | monitoring | photometric |
| Apr 16 | 4:26–11:59 | 31in | +3.590 | 1.046 | 0.158 | 70.1 | 52.7 | R, CN | monitoring | photometric |
| Apr 17 | 4:56–11:58 | 31in | +4.600 | 1.047 | 0.160 | 70.1 | 53.6 | R, CN | monitoring | clouds |
| Apr 18 | 6:29–11:58 | 31in | +5.632 | 1.048 | 0.162 | 70.0 | 54.3 | R, CN | monitoring | clouds |
| Apr 18 | 9:42– 9:53 | DCT | +5.656 | 1.048 | 0.162 | 70.0 | 54.3 | r', CN | snapshot | clouds |
| Apr 19 | 4:22–11:43 | 31in | +6.583 | 1.049 | 0.164 | 69.9 | 54.8 | R, CN | monitoring | photometric |
| Apr 20 | 4:28–11:49 | 31in | +7.587 | 1.050 | 0.166 | 69.7 | 55.3 | R, CN | monitoring | photometric |
| Apr 21 | 5:51–11:03 | 42in | +8.600 | 1.052 | 0.168 | 69.4 | 55.6 | R, CN, OH, C3, C2, BC | imaging | photometric |
| Apr 22 | 4:22–11:00 | 42in | +9.568 | 1.053 | 0.170 | 69.2 | 55.8 | R, CN, OH, C3, C2, BC | imaging | photometric |
| Apr 23 | 4:27–10:17 | 42in | +10.555 | 1.055 | 0.173 | 68.8 | 56.0 | R, CN, C3 | imaging | photometric |
| Apr 24 | 4:46– 9:27 | 42in | +11.544 | 1.057 | 0.175 | 68.4 | 56.0 | R, CN, OH | imaging | intermittent clouds |
| Apr 24 | 5:14–11:51 | 31in | +11.604 | 1.057 | 0.175 | 68.4 | 56.0 | R, CN | monitoring | intermittent clouds |
| Apr 26 | 5:19–10:41 | 42in | +13.581 | 1.061 | 0.180 | 67.5 | 56.0 | R, CN | monitoring | cirrus |
| Apr 26 | 5:11–11:46 | 31in | +13.601 | 1.061 | 0.180 | 67.5 | 56.0 | R, CN | monitoring | cirrus |
| Apr 27 | 5:03–11:54 | 31in | +14.601 | 1.064 | 0.183 | 67.0 | 55.8 | R, CN | monitoring | clouds |
| Apr 28 | 5:03–11:54 | 31in | +15.601 | 1.067 | 0.185 | 66.4 | 55.7 | R, CN | monitoring | clouds |
| Apr 29 | 5:17–11:00 | 31in | +16.587 | 1.069 | 0.188 | 65.9 | 55.5 | R, CN, C3 | imaging | photometric |
| Apr 30 | 5:18–11:01 | 31in | +17.588 | 1.072 | 0.190 | 65.2 | 55.2 | R, CN, C3 | imaging | cirrus |
| May 1 | 5:19–10:29 | 31in | +18.577 | 1.075 | 0.193 | 64.6 | 54.9 | R, CN, C3 | imaging | photometric |
| May 1 | 11:20–11:29 | DCT | +18.723 | 1.076 | 0.193 | 64.5 | 54.9 | r', CN | snapshot | photometric |
| May 2 | 11:12–11:39 | 31in | +19.724 | 1.079 | 0.196 | 63.8 | 54.6 | R, CN | snapshot | photometric |
| May 3 | 5:20–10:29 | 31in | +20.577 | 1.082 | 0.199 | 63.2 | 54.3 | R, CN, C3 | monitoring | cirrus |
| May 4 | 9:58–10:20 | 31in | +21.671 | 1.086 | 0.202 | 62.4 | 53.9 | R, CN, C3 | snapshot | cirrus |
| May 14 | 10:08–10:18 | DCT | +31.673 | 1.130 | 0.231 | 53.9 | 49.3 | r', CN | snapshot | photometric |
| May 17 | 5:01–10:45 | 31in | +34.576 | 1.145 | 0.240 | 51.1 | 47.8 | R, CN | monitoring | intermittent clouds |
| May 20 | 10:47–11:01 | DCT | +37.702 | 1.162 | 0.250 | 48.0 | 45.9 | r', CN | snapshot | photometric |
| May 28 | 10:18–10:32 | DCT | +45.682 | 1.211 | 0.279 | 40.6 | 40.6 | r', CN | snapshot | photometric |
| May 29 | 10:35–10:49 | DCT | +46.694 | 1.218 | 0.283 | 38.9 | 39.8 | r', CN | snapshot | cirrus |
| Jun 3 | 10:16–10:30 | DCT | +51.680 | 1.252 | 0.303 | 33.9 | 35.7 | r', CN | snapshot | clouds |
| Jun 4 | 10:15–10:33 | DCT | +52.681 | 1.259 | 0.308 | 32.9 | 34.7 | r', CN | snapshot | clouds |
| Jun 18 | 10:49–10:58 | DCT | +66.702 | 1.364 | 0.381 | 20.5 | 15.6 | r', CN | snapshot | photometric |
| Jul 2 | 8:45– 9:04 | DCT | +80.619 | 1.477 | 0.482 | 14.2 | 338.1 | r', CN | snapshot | photometric |

[a] All parameters were taken at the midpoint of each night's observations.

[b] Telescope used: DCT = Discovery Channel Telescope, 42in = 42-inch (1.1-m) Hall telescope, 31in = 31-inch (0.8-m) telescope.

[c] Time from perihelion.

[d] Heliocentric distance.

[e] Geocentric distance.

[f] Solar phase angle.

[g] Position angle (P.A.) of the Sun.

Table 2. Matching images used for period determination for 41P/Tuttle-Giacobini-Kresák.

| UT Date (first image) | UT Date (second image) | Mid-time [a] (day) | $\Delta t$ [b] (hr) | Cycles | Period (hr) |
|---|---|---|---|---|---|
| Mar 18  3:09 | Mar 21  2:47 | −24.130 | 71.63 | 3 | 23.878 |
| Mar 19  4:36 | Mar 21  3:49 | −23.579 | 47.22 | 2 | 23.608 |
| Mar 19  4:36 | Mar 21  4:50 | −23.557 | 48.23 | 2 | 24.117 |
| Mar 19 10:16 | Mar 21 10:14 | −23.327 | 47.97 | 2 | 23.983 |
| Mar 20  4:07 | Mar 21  3:49 | −23.089 | 23.70 | 1 | 23.700 |
| Mar 19 10:23 | Mar 22 11:08 | −22.806 | 72.75 | 3 | 24.250 |
| Mar 21  3:29 | Mar 22  3:34 | −22.107 | 24.08 | 1 | 24.083 |
| Mar 21  3:49 | Mar 22  4:06 | −22.089 | 24.28 | 1 | 24.283 |
| Mar 21  8:53 | Mar 22  8:36 | −21.890 | 23.72 | 1 | 23.717 |
| Mar 21  9:13 | Mar 22  9:13 | −21.870 | 24.00 | 1 | 24.000 |
| Mar 21 10:57 | Mar 22 11:08 | −21.794 | 24.18 | 1 | 24.183 |
| Mar 18  3:09 | Mar 27 10:58 | −20.960 | 223.82 | 9 | 24.869 |
| Mar 19  4:36 | Mar 26  9:56 | −20.951 | 173.33 | 7 | 24.762 |
| Mar 21  7:00* | Mar 24 10:11 | −20.896 | 75.18 | 3 | 25.061 |
| Mar 20  4:07 | Mar 25 11:08 | −20.937 | 127.02 | 5 | 25.403 |
| Mar 21  9:36 | Mar 24 11:08 | −20.822 | 73.53 | 3 | 24.511 |
| Mar 22  6:44 | Mar 24  8:25 | −20.439 | 49.68 | 2 | 24.842 |
| Mar 20  4:07 | Mar 26  8:10 | −20.498 | 148.05 | 6 | 24.675 |
| Mar 22  7:59 | Mar 24  9:20 | −20.393 | 49.35 | 2 | 24.675 |
| Mar 22  7:59 | Mar 24 10:11 | −20.375 | 50.20 | 2 | 25.100 |
| Mar 20  4:17 | Mar 26  9:56 | −20.458 | 149.65 | 6 | 24.942 |
| Mar 22  8:55* | Mar 24 10:11 | −20.356 | 49.27 | 2 | 24.633 |
| Mar 22  9:13 | Mar 24 11:08 | −20.330 | 49.92 | 2 | 24.958 |
| Mar 22  3:34 | Mar 25  6:28 | −20.045 | 74.90 | 3 | 24.967 |
| Mar 21  3:29 | Mar 26  9:03 | −19.993 | 125.57 | 5 | 25.113 |
| Mar 21  4:30 | Mar 26  9:25* | −19.964 | 124.92 | 5 | 24.983 |
| Mar 21  4:30 | Mar 26  9:56 | −19.953 | 125.43 | 5 | 25.087 |
| Mar 21  4:50 | Mar 26 10:05* | −19.943 | 125.25 | 5 | 25.050 |
| Mar 21  5:51 | Mar 26 11:10* | −19.900 | 125.32 | 5 | 25.063 |
| Mar 21  6:10 | Mar 26 11:36 | −19.884 | 125.43 | 5 | 25.087 |
| Mar 20  4:07 | Mar 27 11:18 | −19.933 | 175.18 | 7 | 25.026 |
| Mar 22  3:02 | Mar 26  9:03 | −19.503 | 102.02 | 4 | 25.504 |
| Mar 22  3:34 | Mar 26  9:03 | −19.491 | 101.48 | 4 | 25.371 |
| Mar 22  3:34 | Mar 26  9:30* | −19.482 | 101.93 | 4 | 25.483 |
| Mar 22  4:06 | Mar 26  9:50* | −19.464 | 101.73 | 4 | 25.433 |
| Mar 22  4:48 | Mar 26  9:56 | −19.447 | 101.13 | 4 | 25.283 |
| Mar 22  5:31 | Mar 26 10:40* | −19.417 | 101.15 | 4 | 25.288 |
| Mar 24  8:25 | Mar 25  9:48 | −18.875 | 25.38 | 1 | 25.383 |
| Mar 24  8:25 | Mar 25 10:20 | −18.864 | 25.92 | 1 | 25.917 |
| Mar 24  8:25 | Mar 26 12:00* | −18.329 | 51.58 | 2 | 25.792 |
| Mar 25  4:40 | Mar 26  6:29 | −18.022 | 25.82 | 1 | 25.817 |
| Mar 25  5:36 | Mar 26  7:17 | −17.986 | 25.68 | 1 | 25.683 |
| Mar 25  5:36 | Mar 26  8:10 | −17.967 | 26.57 | 1 | 26.567 |
| Mar 25  7:21 | Mar 26  9:56 | −17.894 | 26.58 | 1 | 26.583 |
| Mar 25  4:40 | Mar 27  8:55* | −17.471 | 52.25 | 2 | 26.125 |
| Mar 22 11:08 | Mar 30  3:51 | −17.442 | 184.72 | 7 | 26.388 |
| Mar 25  5:36 | Mar 27  9:50* | −17.432 | 52.23 | 2 | 26.117 |
| Mar 25  5:36 | Mar 27 10:22 | −17.421 | 52.77 | 2 | 26.383 |
| Mar 25  6:28 | Mar 27 10:58 | −17.391 | 52.50 | 2 | 26.250 |
| Mar 25  7:21 | Mar 27 11:39 | −17.358 | 52.30 | 2 | 26.150 |
| Mar 26  4:43 | Mar 27  6:40* | −17.017 | 25.95 | 1 | 25.950 |
| Mar 26  6:29 | Mar 27  8:10* | −16.949 | 25.68 | 1 | 25.683 |
| Mar 26  7:17 | Mar 27  9:00* | −16.915 | 25.72 | 1 | 25.717 |
| Mar 26  7:17 | Mar 27 10:22 | −16.887 | 27.08 | 1 | 27.083 |

Table 2—Continued

| UT Date (first image) | UT Date (second image) | Mid-time[a] (day) | $\Delta t$[b] (hr) | Cycles | Period (hr) |
|---|---|---|---|---|---|
| Mar 26  8:10 | Mar 27  10:00* | −16.875 | 25.83 | 1 | 25.833 |
| Mar 26  9:03 | Mar 27  10:50* | −16.840 | 25.78 | 1 | 25.783 |
| Mar 26  9:56 | Mar 27  11:39  | −16.805 | 25.72 | 1 | 25.717 |
| Mar 27  2:30* | Mar 30  11:54 | −14.954 | 81.40 | 3 | 27.133 |
| Apr 11  6:08 | Apr 17  5:12 | +1.482 | 143.07 | 4 | 35.767 |
| Apr 11  9:32 | Apr 17  11:45 | +1.689 | 146.22 | 4 | 36.554 |
| Apr 11  4:54 | Apr 20  11:36 | +3.090 | 222.70 | 6 | 37.117 |
| Apr 24  8:37 | Apr 26  5:32 | +12.541 | 44.92 | 1 | 44.917 |
| Apr 24  8:37 | Apr 28  5:19 | +13.536 | 92.70 | 2 | 46.350 |
| Apr 26  8:34 | Apr 28  7:39 | +14.584 | 47.08 | 1 | 47.083 |
| Apr 28  7:39 | Apr 30  8:40 | +16.586 | 49.02 | 1 | 49.017 |
| Apr 29  7:36 | May 1  8:40 | +17.585 | 49.07 | 1 | 49.067 |
| Apr 29  5:30 | May 3  10:13 | +18.573 | 100.72 | 2 | 50.358 |
| May 1  6:35 | May 3  10:13 | +19.596 | 51.63 | 1 | 51.633 |

[a]Time at the midpoint of the image pair in days relative to perihelion ($T$ = 2017 April 12.752).

[b]Time in hours between the image pair.

*Estimated optimal match based on interpolation between neighboring images.

Table 3. Rotationally phased period determinations for comet 41P/Tuttle-Giacobini-Kresák.

| Mid-time[a] | ΔT[b] (day) | Period (hr) |
|---|---|---|
| Mar 22.5 | −21.3 | 24.6 |
| Mar 24.5 | −19.3 | 25.3 |
| Mar 27.5 | −16.3 | 26.5 |
| Apr 8.0 | −4.8 | 38.6 |
| Apr 17.5 | +4.7 | 38.6 |
| Apr 20.9 | +8.2 | 41.5 |
| Apr 23.0 | +10.2 | 42.3 |
| Apr 28.5 | +15.7 | 48.0 |

[a] Approximate time at the midpoint of the period determination in 2017.

[b] Time in days relative to perihelion ($T = 2017$ April 12.752) for the mid-time.